\documentclass[12pt]{iopart}
\usepackage{epsfig,amssymb}

\def\spose#1{\hbox to 0pt{#1\hss}}
\def\lesssim{\mathrel{\spose{\lower 3pt\hbox{$\mathchar"218$}}
 \raise 2.0pt\hbox{$\mathchar"13C$}}}
\def\gtrsim{\mathrel{\spose{\lower 3pt\hbox{$\mathchar"218$}}
 \raise 2.0pt\hbox{$\mathchar"13E$}}}

\def\<{\langle}
\def\>{\rangle}

\begin{document}

\title{ 
Universality class of 3D site-diluted and bond-diluted Ising systems
}

\author{Martin Hasenbusch,$^{1}$ 
Francesco Parisen Toldin,$^2$ \\
Andrea Pelissetto,$^3$ 
and Ettore Vicari$\,^1$ } 

\address{$^1$ Dip. Fisica dell'Universit\`a di Pisa and
INFN, I-56127 Pisa, Italy}

\address{$^2$ Scuola Normale Superiore and
INFN, I-56126 Pisa, Italy} 

\address{$^3$
Dip. Fisica dell'Universit\`a di Roma ``La Sapienza" and INFN, \\ 
I-00185 Roma, Italy}

\ead{
Martin.Hasenbusch@df.unipi.it, 
parisen@sns.it,
Andrea.Pelissetto@roma1.infn.it,
Ettore.Vicari@df.unipi.it}

\begin{abstract}

We present a finite-size scaling analysis of high-statistics Monte
Carlo simulations of the three-dimensional randomly site-diluted and bond-diluted Ising
model.  The critical behavior of these systems is affected by
slowly-decaying scaling corrections which make the accurate
determination of their universal asymptotic behavior quite hard,
requiring an effective control of the scaling corrections.  For this
purpose we exploit improved Hamiltonians, for which the leading
scaling corrections are suppressed for any thermodynamic quantity, and
improved observables, for which the leading scaling corrections are
suppressed for any model belonging to the same universality class.

The results of the finite-size scaling analysis provide strong numerical
evidence that phase transitions in three-dimensional randomly site-diluted and
bond-diluted Ising models belong to the same randomly dilute Ising
universality class.  We obtain accurate estimates of the critical exponents,
$\nu=0.683(2)$, $\eta=0.036(1)$, $\alpha=-0.049(6)$, $\gamma=1.341(4)$,
$\beta=0.354(1)$, $\delta=4.792(6)$, and of the leading and next-to-leading
correction-to-scaling exponents, $\omega=0.33(3)$ and $\omega_2=0.82(8)$.
\end{abstract}

\bigskip

\noindent

\maketitle


\section{Introduction}
\label{intro}

The effect of quenched random disorder on the critical behavior of
Ising systems has been much investigated experimentally and theoretically.
Typical physical examples are randomly dilute uniaxial antiferromagnets, for
instance, Fe$_{p}$Zn$_{1-p}$F${}_2$ and Mn$_{p}$Zn$_{1-p}$F${}_2$, obtained by
mixing a uniaxial antiferromagnet with a nonmagnetic material.  Experiments
(see, e.g., Ref.~\cite{Belanger-00} for a review) find that, for sufficiently
low impurity concentration $1-p$, these systems undergo a second-order phase
transition at $T_c(p) < T_c(p=1)$. The critical behavior appears approximately
independent of the impurity concentration, but definitely different from the
one of the pure system. These results support the existence
of a random Ising universality class which differs from the one of pure Ising
systems.\footnote{Unixial antiferromagnets undergo a phase transition also in the 
presence of a magnetic field $H$. In the absence of dilution, for small $H$ 
the transition is in the Ising universality class: the magnetic field does not change
the critical behavior. In the presence of dilution instead, $H$ is relevant and 
the critical behavior for $H\not=0$ belongs to the random-field Ising universality
class~\cite{cross-to-randomfield}. The crossover occurring for $H\to 0$
is studied in Ref.~\cite{CPV-crossover}. }

A simple lattice model for dilute Ising systems is provided by the
three-dimensional randomly site-diluted Ising model (RSIM) with Hamiltonian
\begin{equation}
{\cal H}_{s} = -J\,\sum_{<xy>}  \rho_x \,\rho_y \; \sigma_x \sigma_y,
\label{Hs}
\end{equation}
where the sum is extended over all nearest-neighbor sites of a simple cubic
lattice, $\sigma_x$ are Ising spin variables, and $\rho_x$ are uncorrelated
quenched random variables, which are equal to one with probability $p$ (the
spin concentration) and zero with probability $1-p$ (the impurity
concentration).  Another related lattice model is the randomly bond-diluted
Ising model (RBIM) with Hamiltonian
\begin{equation}
{\cal H}_{b} = -J\,\sum_{<xy>} j_{xy} \; \sigma_x \sigma_y,
\label{Hb}
\end{equation}
where the bond variables $j_{xy}$ are uncorrelated quenched random variables,
which are equal to one with probability $p$ and zero with probability $1-p$.
Note that the RSIM can be seen as a RBIM with bond variables $\hat{\hbox{\em \j}}_{xy} =
\rho_x \rho_y$. But in this case the bond variables are correlated. For
example, the connected average of the bond variables along a plaquette does not
vanish as in the case of uncorrelated bond variables: indeed 
(note that $\overline{\hat{\hbox{\em \j}}_{xy}} = p^2$)
\begin{equation}
\overline{\prod_\square (\hat{\hbox{\em \j}}_{xy} - 
  \overline{\hat{\hbox{\em \j}}_{xy}})} = p^4 - 4 p^6 + 4 p^7 - p^8.
\end{equation}
Above the percolation threshold of the spins, 
these models undergo a phase transition between a disordered and a
ferromagnetic phase.  Its nature
has been the object of many theoretical studies, see, e.g.,
Refs.~\cite{Aharony-76,Stinchcombe-83,PV-02,FHY-03,JBCBH-05} for reviews.  A
natural scenario is that the critical behavior of the RSIM and the RBIM, at any value
of $p$ above the spin percolation threshold, belongs to the same
universality class, which we will call randomly dilute Ising (RDIs) universality class.

\begin{figure}[tb]
\vspace{1cm}
\centerline{\psfig{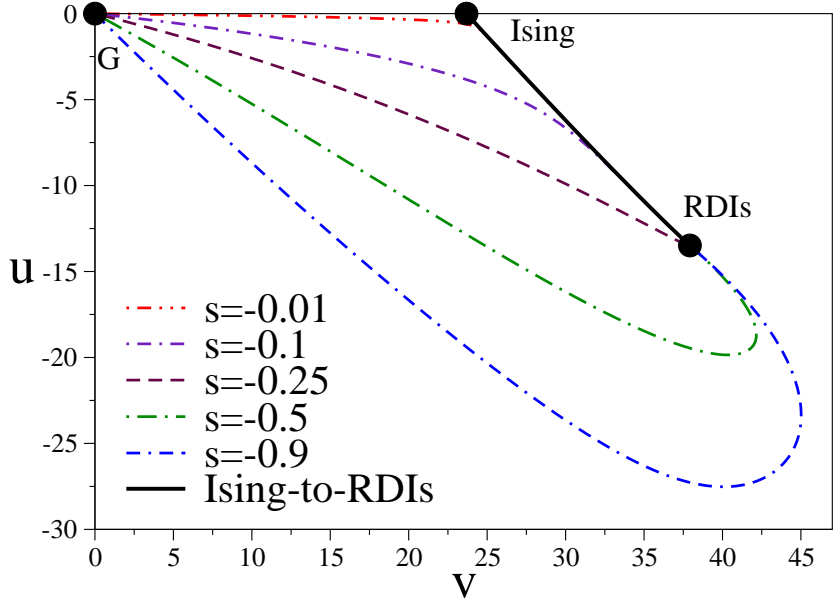}}
\vspace{2mm}
\caption{
  RG trajectories in the renormalized coupling $(u,v)$ plane starting from the
  Gaussian FP, for several
  values of the ratio $s\equiv u_0/v_0$ 
  (from Ref.~\cite{CPPV-04}). We also report the RG trajectory connecting 
  the Ising FP with the RDIs FP, relevant for the Ising-to-RDIs crossover. }
\label{figtraj}
\end{figure}

The RDIs universality class can be investigated by field-theoretical (FT)
methods, starting from the Landau-Ginzburg-Wilson $\phi^4$ Hamiltonian
\begin{equation}
H_{\phi^4} =  \int d^d x \Bigl\{ {1\over 2} \sum_{i=1}^{N}
      \left[ (\partial_\mu \phi_i)^2 +  r \phi_i^2 \right]  
+{1\over 4!} \sum_{i,j=1}^N \left( u_0 + v_0 \delta_{ij} \right)
\phi^2_i \phi^2_j  \Bigr\}, 
\label{Hphi4rim}
\end{equation}
where $\phi_i$ is an $N$-component field.  By using the standard replica
trick, it can be shown that, for $u_0<0$ and in the limit $N\to 0$, this 
model corresponds to a system with quenched disorder effectively coupled to
the energy density.  Fig.~\ref{figtraj} shows the renormalization-group (RG)
flow \cite{CPPV-04} in the $u,v$ plane, where
$u,v$ are the renormalized couplings associated with the Hamiltonian
parameters $u_0$ and $v_0$. The RG flow has a stable fixed point (FP) in the
region $u<0$, which attracts systems with $-1\lesssim u_0/v_0 < 0$.  The
standard Ising FP at $u=0$ is unstable, with a crossover exponent \cite{Aharony-76}
$\phi=\alpha_{\rm Is}$, where \cite{CPRV-02}
$\alpha_{\rm Is}=0.1096(5)$ is
the specific-heat exponent of the Ising universality class.
The stable RDIs FP determines the critical behavior at the
phase transition between the disordered and the ferromagnetic phase that 
occurs in RDIs systems.
Therefore, it is expected to determine the critical behavior of the RSIM and of the RBIM
above the spin percolation threshold.  The critical exponents at the RDIs FP
have been computed perturbatively to six loops in the three-dimensional 
massive zero-momentum scheme \cite{PV-00,PS-00}. Even though the 
perturbative series are not Borel summable \cite{BMMRY-87,McKane-94,AMR-00},
appropriate resummations provide quite accurate results \cite{PV-00}:
$\nu=0.678(10)$, $\alpha=-0.034(30)$, $\eta=0.030(3)$, $\gamma=1.330(17)$,
$\beta=0.349(5)$.  Moreover, the leading scaling corrections are 
characterized by a small exponent $\omega=0.25(10)$, 
which is much smaller than that occurring 
in pure Ising systems, $\omega\approx 0.8$ \cite{GZ-98,PV-02}. 
Experiments find \cite{Belanger-00} $\nu=0.69(1)$,
$\alpha=-0.10(2)$, and $\beta=0.350(9)$, which are in reasonable agreement
with the FT estimates (there is only a small discrepancy for $\alpha$).
We also mention that approximate
expressions for the RDIs critical equation of state have been reported in
Ref.~\cite{CDPV-03}.

Monte Carlo (MC) simulations of the RSIM and the RBIM have long been
inconclusive in settling the question of the critical behavior of these
models. While the measured critical exponents were definitely different from
the Ising ones, results apparently depended on the spin concentration, in
disagreement with RG theory.  The question was clarified in
Ref.~\cite{BFMMPR-98}, where the apparent violations of universality were
explained by the presence of large concentration-dependent scaling
corrections, which decay very slowly because of the small value of the
exponent $\omega$, $\omega = 0.37(6)$. Only if they are properly taken into
account, the numerical estimates of the critical exponents of the RSIM become
dilution-independent as expected. Ref.~\cite{BFMMPR-98} reported the estimates
$\nu=0.6837(53)$ and $\eta=0.0374(45)$, which are in agreement with the FT
results. These results were later confirmed by MC simulations \cite{CMPV-03}
of the RSIM at $p=0.8$, which is the value of $p$ where the leading scaling
corrections appear suppressed \cite{BFMMPR-98,Hukushima-00}.  A FSS analysis
of the data up to lattice size $L=256$ gave $\nu = 0.683(3)$ and $\eta =
0.035(2)$. On the other hand, results for the RBIM have been less
satisfactory.  Recent works, based on FSS analyses~\cite{BCBJ-04} of MC data
up to $L=96$ and high-temperature expansions~\cite{HJ-06}, have apparently
found asymptotic power-law behaviors that are quite dependent on the spin
concentration.  Such results may again be explained by the presence of
sizeable and spin-concentration dependent scaling corrections.

The RSIM and the RBIM, and in general systems which are supposed to belong to
the RDIs universality class, are examples of models in which the presence of
slowly-decaying scaling corrections makes the determination of the asymptotic
critical behavior quite difficult. In these cases, the universal critical
behavior can be reliably determined only if scaling corrections are kept under
control in the numerical analyses.  For example, the Wegner expansion of the
magnetic susceptibility $\chi$ is generally given by~\cite{Wegner-76}
\begin{eqnarray}
\chi &=& C t^{-\gamma} 
\left( 1 + a_{0,1} t + a_{0,2}t^2 + \ldots + a_{1,1}
t^{\Delta} + a_{1,2} t^{2\Delta} + \ldots \right.
\nonumber \\ 
&& \left. \qquad + b_{1,1} t^{1+\Delta} +
b_{1,2} t^{1+2\Delta} +
\ldots + a_{2,1} t^{\Delta_2} + \ldots \right),
\label{chiwexp}
\end{eqnarray}
where $t\equiv 1 - \beta/\beta_c$ is the reduced temperature. We have
neglected additional terms due to other irrelevant operators and terms due to
the analytic background present in the free energy \cite{AF-83,SS-00,CHPV-02}.
In the case of the three-dimensional RDIs universality class we have
\cite{BFMMPR-98,PV-00} $\Delta =\omega \nu\approx 0.2$, which is very small,
and \cite{CMPV-03} $\Delta_2=\omega_2\nu\approx 0.5$.  Analogously, the
finite-size scaling (FSS) behavior at criticality is given by
\cite{SS-00,CHPV-06}
\begin{equation}
\chi = c L^{2-\eta} 
\left( 1 + a_{1,1}
L^{-\omega} + a_{1,2} L^{-2\omega} + \ldots 
+ a_{2,1} L^{-\omega_2} + \ldots \right),
\label{chifss}
\end{equation}
where $\omega\approx 0.3$ and $\omega_2\approx 0.8$ for the RDIs universality
class.

The main purposes of this paper are the following:

\begin{itemize}
\item[(i)] We wish to improve the numerical estimates of 
the critical exponents associated with the asymptotic behavior and with the 
leading scaling corrections.
\item[(ii)]
We wish to provide robust evidence that
the critical behaviors of the RSIM and of the RBIM belong to the same RDIs
universality class, independently of the impurity concentration.
\end{itemize}

For these purposes, we perform a high-statistics MC simulation of the RSIM for
$p=0.8$ and $p=0.65$, and of the RBIM for $p=0.7$ and $p=0.55$, for lattice
sizes up to $L=192$. The critical behavior is obtained by a careful FSS
analysis, in which a RG invariant quantity (we shall use $R_\xi\equiv \xi/L$)
is kept fixed. This method has significant advantages
\cite{Has-99,HPV-05,CHPV-06} with respect to more standard approaches, and, in
particular, it does not require a precise estimate of $\beta_c$. Our main
results can be summarized as follows.

\begin{itemize}
\item 
 We obtain accurate estimates of the critical exponents: $\nu=0.683(2)$
  and $\eta=0.036(1)$. Then, using the scaling and hyperscaling relations
  $\alpha=2-3\nu$, $\gamma=(2-\eta)\nu$, $\beta=\nu (1+\eta)/2$ and
  $\delta=(5-\eta)/(1+\eta)$, we obtain $\alpha=-0.049(6)$, $\gamma=1.341(4)$,
  $\beta=0.354(1)$, and $\delta=4.792(6)$.

\item
We obtain accurate estimates of the exponents associated with the
leading scaling corrections: $\omega =0.33(3)$ and $\omega_2=0.82(8)$.
Correspondingly, we have  $\Delta =\omega \nu = 0.22(2)$ and $\Delta_2 =
\omega_2\nu=0.56(5)$.

\item For both the RSIM and the RBIM we estimate the value $p^*$ at which 
  the leading scaling corrections associated with the
  exponent $\omega$ vanish for all quantities. 
  For the RSIM, we find $p^*=0.800(5)$.  This result significantly
  strengthens the evidence that the RSIM with $p=0.8$ is 
  improved, as already suggested by earlier calculations
  \cite{BFMMPR-98,CMPV-03}. For the RBIM, we find $p^* = 0.54(2)$.

\item 
We provide strong evidence that the transitions in the RSIM and in the RBIM 
belong to the same universality class.  For this purpose, the knowledge of
the leading and next-to-leading scaling correction exponents is
essential. We also make use of improved observables characterized by
the (approximate) absence of the leading scaling correction for any system
belonging to the RDIs universality class.

\end{itemize}

The paper is organized as follows. In Sec.~\ref{definitions} we report the
definitions of the quantities which are considered in the paper.  In
Sec.~\ref{FSS} we summarize some basic FSS results which are needed for the
analysis of the MC data.  In Sec.~\ref{MC} we give some details of the MC
simulations.  Sec.~\ref{FSSRSIM} describes the FSS analyses of the MC data of
the RSIM at $p=0.8$ and $p=0.65$ at fixed $R_\xi$, which lead to the best
estimates of the critical exponents. FSS analyses at fixed $\beta$ of the RSIM
at $p=0.8$ are presented in Sec.~\ref{FSSfixedbeta}. Finally, in
Sec.~\ref{FSSRBIM} we analyze the data for the RBIM at $p=0.55$ and $p=0.7$,
and show that the RBIM belongs to the same universality class as the RSIM. In
\ref{omega2} we determine the next-to-leading correction-to-scaling exponent
$\omega_2$ by a reanalysis of the FT six-loop perturbative expansions reported
in Ref.~\cite{CPV-00}.  In \ref{appendix_bias} we discuss the problem of the
bias in MC calculations of disorder averages of combinations of thermal
averages.

\section{Notations} \label{definitions}

We consider Hamiltonians (\ref{Hs}) and (\ref{Hb}) with $J = 1$ on
a finite simple-cubic lattice $L^3$ with periodic boundary conditions.
In the case of the RSIM, given a quantity ${\cal O}$ depending on the
spins $\{\sigma\}$ and on the random variables $\{\rho\}$, we define the
thermal average at fixed distribution $\{\rho\}$ as
\begin{equation}
\langle {\cal O} \rangle (\beta,\{\rho\}) \equiv 
    {1\over Z(\{\rho\})} \sum_{\{\sigma\}} {\cal O} e^{-\beta {\cal H}[\sigma,\rho]},
\end{equation}
where $Z(\{\rho\})$ is the sample partition function. Then, we average over
the random dilution considering
\begin{equation}
\overline{\langle {\cal O} \rangle} (\beta) = 
   \int [d\rho] \langle {\cal O} \rangle (\beta,\{\rho\}) ,
\end{equation}
where 
\begin{equation}
[d\rho] = \prod_x [p \delta(\rho_x - 1) + (1-p) \delta(\rho_x) ].
\end{equation}
Analogous formulas can be written for the RBIM, taking
\begin{equation}
[dj] = \prod_{< xy >}  [p \delta(j_{xy} - 1) + (1-p) \delta(j_{xy}) ].
\end{equation}
We define the two-point correlation function
\begin{eqnarray}
&G(x) \equiv \overline{\langle \rho_0 \sigma_0 \,\rho_x \sigma_x \rangle} 
\qquad & \hbox{(RSIM)},  \nonumber  \\
&G(x) \equiv \overline{\langle \sigma_0 \,\sigma_x \rangle} 
\qquad & \hbox{(RBIM)},
\label{twop}
\end{eqnarray}
the corresponding susceptibility $\chi$,
\begin{equation}
\chi \equiv \sum_x G(x),
\end{equation}
and the correlation length $\xi$,
\begin{equation}
\xi^2 \equiv {\widetilde{G}(0) - \widetilde{G}(q_{\rm min}) \over 
          \hat{q}_{\rm min}^2 \hat{G}(q_{\rm min}) },
\end{equation}
where $q_{\rm min} \equiv (2\pi/L,0,0)$, $\hat{q} \equiv 2 \sin q/2$,
and $\widetilde{G}(q)$ is the Fourier transform of $G(x)$.


We also consider  quantities
that are invariant under RG transformations in the critical limit.
We call them phenomenological couplings and 
generically refer to them by using the symbol $R$.  Beside the ratio
\begin{equation}
R_\xi = \xi/L,
\label{rxi}
\end{equation}
we consider the quartic cumulants $U_4$ and $U_{22}$ defined by
\begin{eqnarray}
&& U_{4}  \equiv { \overline{\mu_4}\over \overline{\mu_2}^{2}}, 
\qquad 
U_{22} \equiv  {\overline{\mu_2^2}-\overline{\mu_2}^2 \over \overline{\mu_2}^2},
\label{cumulants}
\end{eqnarray}
where
\begin{eqnarray}
&\mu_{k} \equiv \langle \; ( \sum_x \rho_x \sigma_x\; )^k \rangle 
\quad & \quad {\rm (RSIM)},  \\
&\mu_{k} \equiv \langle \; ( \sum_x \sigma_x\; )^k \rangle 
\quad & \quad {\rm (RBIM)}.  
\nonumber
\end{eqnarray}
We also define their difference
\begin{equation}
U_d \equiv U_4 - U_{22}.
\label{udiff}
\end{equation}
Finally, we consider the derivative of the phenomenological
couplings $R$ with respect to the inverse temperature, i.e.,
\begin{equation}
R' \equiv { \partial R\over \partial \beta},
\label{sdef}
\end{equation}
which allows one to determine the critical exponent $\nu$.

\section{Finite-size scaling}
\label{FSS}

\subsection{General results}
\label{FSS-general}

In this section we summarize some basic results concerning FSS, which will be
used in the FSS analysis.  We consider a generic RDIs model in the presence of
a constant magnetic field $H$ and in a finite volume of linear size $L$, and
the disorder-averaged free-energy density
\begin{equation}
{\cal F}(\beta,H,L) = {1\over L^d} \overline{\ln Z(\beta,H,L)},
\end{equation}
where $d$ is the space dimension ($d=3$ in our specific case).  In analogy
with what is found in systems without disorder (see, e.g.,
Refs.~\cite{SS-00,Privman-90}), we expect the free energy to be the sum of a
regular part ${\cal F}_{\rm reg}(\beta,H,L)$ and of a singular part ${\cal
  F}_{\rm sing}(\beta,H,L)$.  The regular part is expected to depend on $L$
only through exponentially small terms, while the singular part encodes the
critical behavior.  The scaling behavior of the latter is expected to be:
\begin{eqnarray}
{\cal F}_{\rm sing} (u_t, u_h, \{u_i\},L)  =
 L^{-d}
{\cal F}_{\rm sing}( L^{y_t} u_t, L^{y_h} u_h, \{L^{y_i} u_i\}),
\label{FscalL}
\end{eqnarray}
where $u_t\equiv u_1$, $u_h\equiv u_2$, $\{u_i\}$ with $i\geq 3$ are the
scaling fields associated respectively with the reduced temperature $t$
($u_t\sim t$), the magnetic field $H$ ($u_h\sim H$), and the other irrelevant
perturbations with $y_i<0$.  They are analytic functions of the Hamiltonian
parameters---in particular, of $t$ and $H$---and are expected not to depend on
the linear size $L$ \cite{SS-00}.  Since $u_t$ and $u_h$ are assumed to be the
only relevant scaling fields, $y_i< 0$ for $i\ge 3$. Thus, in the
infinite-volume limit and for any $t$, the arguments $L^{y_i} u_i$ go to zero.
One may thus expand ${\cal F}_{\rm sing}$ with respect to $L^{y_i} u_i$
obtaining all scaling corrections.  The RG dimensions of the relevant scaling
fields $u_t$ and $u_h$ are related to the standard exponents $\nu$ and $\eta$
by $y_t=1/\nu$ and $y_h=(d+2-\eta)/2$. The correction-to-scaling exponents
$\omega$ and $\omega_2$ introduced in Sec.~\ref{intro} are related to the RG
dimensions of the two leading irrelevant scaling fields: $\omega=-y_3$ and
$\omega_2=-y_4$.

The scaling behavior of zero-momentum thermodynamic quantities can be obtained
by performing appropriate derivatives of Eq.~(\ref{FscalL}) with respect to
$t$ and $H$.  For instance, for the susceptibility at $H = 0$ we obtain
\begin{equation}
\chi(\beta,L) =
k_\chi L^{2-\eta} \left[
\chi_0(u_t L^{y_t}) + 
\sum_{i\ge 3}^\infty \sum_k \chi_{i,k}(u_t L^{y_t}) \, u_i^k L^{k y_i} \right] 
+ \chi_{\rm reg}(\beta),
\label{chiexp_1}
\end{equation}
where we have neglected terms scaling as $L^{y_t - 2y_h}$, $L^{y_i - y_h}$,
and $L^{y_i + y_j - 2y_h}$, which arise from the $H$ dependence of $u_t$ and
$u_i$ (see, e.g., Refs.~\cite{AF-83,CHPV-00} for a discussion in the
infinite-volume limit).  The functions $\chi_0(z)$ and $\chi_{i,k}(z)$ are
smooth, finite for $z\to0$, and universal once one chooses a specific
normalization condition (which must be independent of the Hamiltonian
parameters) for the scaling fields and for the susceptibility (which amounts
to properly choosing the model-dependent constant $k_\chi$).  The function
$\chi_{\rm reg}(\beta)$ represents the contribution of the regular part of the
free-energy density and is $L$ independent (apart from exponentially small
terms).  For $t\to 0$ we have $u_t(t=0) = 0$, while, generically, we expect
$u_i(t=0)\not=0$. Expanding Eq.~(\ref{chiexp_1}) for $L\to\infty$, one obtains
Eq.~(\ref{chifss}).

Analogous formulae hold for the $2n$-point susceptibilities. They allow us to
derive the scaling behavior of the quartic cumulant $R = U_4$.  We obtain in
the FSS limit
\begin{equation}
R(\beta,L) =
r_0(u_t L^{y_t}) + 
\sum_{i\ge 3}^\infty \sum_k r_{i,k}(u_t L^{y_t}) \, u_i^k L^{k y_i}  + 
r_{\rm reg}(\beta),
\label{Rexp_1}
\end{equation}
where again several irrelevant terms have been neglected.  As before, the
functions $r_0(z)$ and $r_{i,k}(z)$ are smooth, finite for $z\to0$, and
universal.  The function $r_{\rm reg}(\beta)$ is due to the regular part of
the free energy and gives rise to scaling corrections of order $L^{\eta-2}$.
For $t\to 0$, writing $u_t = c_t t + O(t^2)$, we can further expand Eq.\ 
(\ref{Rexp_1}), obtaining
\begin{eqnarray}
&& \hspace{-1cm} 
R(\beta,L) =
R^* + {r}_0'(0) c_t \, t L^{y_t}
  + r_{3,0}(0) u_3(t=0) \, L^{y_3}
  + r_{3,1}(0) u_3(t=0)^2 \, L^{2 y_3}
\nonumber \\
  && + r_{4,0}(0) u_4(t=0) \, L^{y_4}
     + O(t^2 L^{2 y_t}, L^{3 y_3}, t L^{y_t+y_3},L^{2 y_4},L^{y_5}),
\label{expandR}
\end{eqnarray}
where $R^*\equiv r_0(0)$ and we have not written explicitly the corrections
due to $r_{\rm reg}(\beta)$.  Note that no analytic $1/L$ corrections are
expected \cite{SS-00,CHPV-06}.

The scaling behavior of $U_{22}$ can be derived analogously.  In this case,
one should start from the two-replica free-energy density (see, e.g.,
Ref.~\cite{CDPV-03}, App.~B)
\begin{equation}
{\cal F}_2(\beta,H_1,H_2,L) = {1\over L^d}
   \overline{\ln Z(\beta,H_1,L)\ln Z(\beta,H_2,L)}
\end{equation}
and assume, as before, that ${\cal F}_2(\beta,H_1,H_2,L)$ is the sum
of a regular part and of a singular part. The latter should scale as 
\begin{eqnarray}
{\cal F}_{2,\rm sing} (u_t, u_{h_1}, u_{h_2}, \{u_i\},L)  =
 L^{-d}
{\cal F}_{\rm sing}( L^{y_t} u_t, L^{y_h} u_{h_1}, L^{y_h} u_{h_2}, 
         \{L^{y_i} u_i\}),
\label{FscalL2}
\end{eqnarray}
where we have now two magnetic scaling fields such that $u_{h_1} \sim H_1$ and
$u_{h_2} \sim H_2$. Taking the appropriate derivatives, one can verify that
Eq.~(\ref{Rexp_1}) holds for $U_{22}$ too.

The discussion we have presented applies only to zero-momentum quantities.  In
order to derive the scaling behavior of the correlation length, we should also
consider quantities that are defined in terms of the field variables at
nonzero momentum.  They can be derived from a free-energy density associated
with a model in which the magnetic field is site dependent. The general
analysis by Wegner \cite{Wegner-76} indicates that Eq.~(\ref{FscalL2}) should
still hold (at least in systems without disorder; we assume here that the same
holds for quenched averages): the presence of a site-dependent $H$ only
modifies the scaling fields that become functionals of $H$. For these reasons,
we conjecture that also $\xi$ has an expansion similar to (\ref{Rexp_1}): in
particular, we expect corrections proportional to $u_3^k L^{-k \omega}$ and
$u_4^k L^{-k\omega_2}$. The momentum dependence of the scaling fields will
give rise to additional terms, the leading one being proportional to $L^{-2}$
(for a discussion in the two-dimensional Ising model, see
Ref.~\cite{CCCPV-00}, p.~8161). Additional $1/L^2$ corrections arise from our
particular definition of $\xi$ \cite{PV-98}.  The analyses we shall report
below confirm this conjecture.

The scaling fields $u_i$ depend on the model. Thus, if one considers families
of models that depend on an irrelevant parameter, by a proper tuning one can
find Hamiltonians for which $u_3(t = 0) = 0$. In this case, in the FSS limit
$t\to 0$, $L\to \infty$ at fixed $t L^{y_t}$, all corrections proportional to
$L^{k y_3} = L^{-k \omega}$ vanish.  Note that, since the scaling field
depends only on the model, such a cancellation occurs for any quantity one
considers. In all cases the leading correction-to-scaling term behaves as
$L^{-\omega_2}$.  Models such that $u_3(t = 0) = 0$ will be called {\em
  improved} models.  Since the leading scaling correction vanishes, one should
observe a faster approach to the scaling limit.

In this paper we shall also consider {\em improved} observables.  An improved
phenomenological coupling is such that $r_{3,0}(0) = 0$.  As a consequence, in
the FSS limit at fixed $u_t L^{y_t} = 0$ it does not show leading scaling
corrections proportional to $L^{-\omega}$.  Note that this cancellation is
only observed on the line in the $(t,L)$ plane such that $t L^{y_t} = 0$, but
not in the generic FSS limit.  As a consequence, if $R$ is improved, its
derivative $R'$ is not.  Note also that, while in improved models all
corrections proportional to $L^{-k \omega}$ vanish, here only the leading one
vanishes: corrections proportional to $L^{-2\omega}$ are still present.
Improved observables satisfy a very important property. Since the function
$r_{3,0}(z)$ is universal, the cancellation occurs in any model.\footnote{In
  principle one could define an improved observable along any line with fixed
  $t L^{y_t}$.  Such observables would be improved in any model on the same
  line $u_t L^{y_t}$.  However, these observables are not very useful in
  practice.  Suppose indeed that one determines an improved observable in a
  model (call it model $A$) along the line such that $t L^{y_t} = k_A$. If
  $u_t = c_A t$, this means that improvement is observed on the line $u_t
  L^{y_t} = k_A c_A$.  Then consider a second model (model $B$). In model $B$
  the observable is improved on the same line $u_t L^{y_t} = k_A c_A$. Now
  $u_t = c_B t$ with $c_B\not=c_A$. Thus, in model $B$ improvement is observed
  on the line $t L^{y_t} = k_A c_A/c_B$. Thus, in order to use the improved
  observable in model $B$ one should determine the model-dependent ratio
  $c_A/c_B$, which is quite inconvenient. No such difficulty arises for $k_A =
  0$. All these problems are avoided by using FSS at fixed phenomenological
  couplings, see next subsection.}

The thermal RG exponent $y_t=1/\nu$ is usually computed from the
FSS of the derivative $R'$ of a phenomenological coupling $R$ with
respect to $\beta$ at $\beta_c$.  Using Eq.\ (\ref{Rexp_1}) one
obtains
\begin{eqnarray}
&&
\hspace{-2.3truecm}
\left. R' \right |_{\beta_c} =
s_0 L^{y_t} \left[ 1 + a_3 u_3(t=0) L^{y_3} + 
 a_4 u_4(t=0) L^{y_4} + 
 b_3 {du_3\over dt}(t=0)\,L^{y_3-y_t} + \ldots \right],
\label{scaling-derivative-R}
\end{eqnarray}
where $s_0$, $a_3$, $a_4$ are constants. 
The leading correction scales as $L^{y_3}=L^{-\omega}$.
In improved models, in which $u_3(t=0) = 0$,
the leading correction is of order $L^{y_4} = L^{-\omega_2}$.
Note that corrections
proportional to $L^{y_3-y_t} = L^{-\omega - 1/\nu}$ are still present 
even if the model is improved.

\subsection{Finite-size scaling at a fixed phenomenological coupling} 
\label{sec3.2}

Instead of computing the various quantities at fixed Hamiltonian
parameters, one may study FSS keeping a phenomenological coupling
$R$ fixed at a given value $R_{f}$ \cite{Has-99}.  This means that,
for each $L$, one considers $\beta_f(L)$ such that
\begin{equation}
R(\beta=\beta_f(L),L) = R_{f}.
\label{rcbeta}
\end{equation}
All interesting thermodynamic quantities are then computed at $\beta =
\beta_f(L)$.  The pseudocritical temperature $\beta_f(L)$ converges to
$\beta_c$ as $L\to \infty$.  The value $R_{f}$ can be specified at will, as
long as $R_f$ is taken between the high- and low-temperature fixed-point
values of $R$.  The choice $R_{f} = R^*$ (where $R^*$ is the
critical-point value) improves the convergence of $\beta_f$ to $\beta_c$ for
$L\to\infty$;  indeed~\cite{Has-99,CHPRV-01}
$\beta_f-\beta_c=O(L^{-1/\nu})$ for generic values of $R_f$, while
$\beta_f-\beta_c=O(L^{-1/\nu-\omega})$ for $R_f=R^*$.  This method has several
advantages. First, no precise knowledge of $\beta_c$ is needed. Secondly, for
some observables, the statistical error at fixed $R_f$ is smaller 
than that  at fixed $\beta = \beta_c$.

If $\beta_f$ is determined as in Eq.~(\ref{rcbeta}), one may consider
the value of other phenomenological couplings $R_\alpha$ at $\beta_f$, 
defining
\begin{equation}
\bar{R}_\alpha(L) \equiv R_\alpha[\beta_f(L),L].
\label{barr}
\end{equation}
The large-$L$ limit of
$\bar{R}_\alpha$ is universal but depends on $R_{f}$ (it differs from the
critical value $R_\alpha^*$, unless $R_{f}=R^*$). Indeed, neglecting scaling
corrections in Eq.~(\ref{Rexp_1}), we have $R_\alpha = r_{\alpha;0}(u_t
L^{y_t})$ and $R = r_{0}(u_t L^{y_t})$, where $r_{\alpha;0}(z)$ and $r_{0}(z)$
are universal functions.  Fixing $R = R_{f}$ corresponds to fixing a
particular trajectory in the $t,L$ plane given by $u_t L^{y_t} = z_f$, where
$z_f$ is the solution of the equation $R_{f} = r_{0}(z_f)$. Along this
trajectory $R_\alpha = \bar{R}_\alpha = r_{\alpha;0}(z_f)$, which shows the
universality of $\bar{R}_\alpha$.

We can define improved observables also when considering FSS at fixed $R_f$.
Such observables show a faster convergence since the corrections to the 
scaling limit scale as $L^{-\omega_2}$ and $L^{-2\omega}$. Moreover, at variance
with the case discussed in Sec.~\ref{FSS-general} where it was practical
to define an improved observable only on the line $t L^{y_t} = 0$, 
here one can choose any value for $R_f$. If one determines an improved 
observable in a given model for a chosen value $R_f$, this observable 
is improved in any other model at the same value $R_f$.

\section{Monte Carlo simulations}
\label{MC}

We performed MC simulations of the RSIM at $p=0.8$ and $p=0.65$ and of the 
RBIM at $p=0.55$ and $p=0.7$ close to the critical temperature 
on cubic lattices of size $L^3$,
with $L=8$, $12$, $16$, $24$, $32$, $48$, $64$, $96$, $128$. For the 
RSIM at $p=0.8$ we also performed simulations with $L = 192$.
For each lattice size, we collected $N_s$ disorder samples,
$N_s$ varying between $10^4$ and $6.4\times 10^5$.
In Tables \ref{tablerun0.8_1},  \ref{tablerun0.8_2}, \ref{tablerun0.65},
\ref{tablerun0.55}, and \ref{tablerun0.7} we report some details of the MC
simulations and the results at fixed $R_\xi\equiv \xi/L = 0.5943$.

In the simulations we used a combination of Metropolis, Swendsen-Wang
(SW) cluster \cite{SW-87}, and Wolff single-cluster \cite{Wolff-89}
updates, to achieve an effective thermalization of short- and
long-range modes.  For each disorder sample, we started from a
random spin configuration, then we typically performed 300 
thermalization steps,
each step consisting in 1 SW update, 1 Metropolis
update, and $L$ single-cluster updates.  Then, we
typically  performed 400 measures for lattice sizes $L
\le 64$ and 600 measures for larger lattices. Between two measurements we
usually performed 1 SW update and $2L$ single-cluster Wolff updates.  
We did several tests of thermalization, performing some runs with
a larger number of thermalization steps; to test the independence of the 
results on $N_m$, we also did few runs with a larger number of measures per 
disorder configuration.

In the determination of the averages over disorder one should take 
care of the bias that occurs because of the finite number of measures 
at fixed disorder \cite{BFMMPR-98-b}. A bias correction should be introduced 
whenever one considers the disorder average of combinations of thermal averages
and, in particular, whenever a reweighting of the data is performed. 
Details are reported in \ref{appendix_bias}.  Errors are computed from the 
sample-to-sample fluctuations and are determined by using the jackknife method.

The MC simulations that we present here
took approximately 10 CPU years of a workstation equipped
with an AMD Opteron Processor 246 (2 GHz clock frequency).

\begin{table}
\caption{
Run parameters and estimates of $\beta_f$ and $\chi$ at fixed $R_\xi= 0.5943$
for the RSIM at $p=0.8$.
$N_s$ is the number of disorder samples divided by $1000$.
}
\label{tablerun0.8_1}
\footnotesize
\begin{center}
\begin{tabular}{@{}rrlll}
\hline
\multicolumn{1}{c}{$L$}&
\multicolumn{1}{c}{$N_s$}&
\multicolumn{1}{c}{$\beta_{\rm run}$}&
\multicolumn{1}{c}{$\beta_f$}&
\multicolumn{1}{c}{$\chi$} \\
\hline 
8  & 200 & 0.285744 & 0.286020(24) & 63.391(9)     \\
12 & 200 & 0.285744 & 0.285851(14) & 141.581(19)   \\
16 & 200 & 0.285744 & 0.285800(6) & 249.910(25)   \\
24 & 100 & 0.285744 & 0.285765(6) & 555.67(7)   \\
32 & 100 & 0.285744 & 0.285751(4) & 979.36(13)   \\
   & 100 & 0.285761 & 0.285753(4) & 979.27(12)   \\
48 &  60 & 0.285744 & 0.2857515(24) & 2173.7(3)   \\
   & 106 & 0.285748 & 0.2857459(18) & 2173.8(3)   \\
   &  60 & 0.285751 & 0.285743(3) & 2174.0(3)   \\
64 & 60 & 0.285742 & 0.2857478(18) & 3827.1(6)   \\
   & 63 & 0.285744 & 0.2857443(17) & 3827.4(6)   \\
96 & 30 & 0.285744 & 0.2857441(13) & 8491.6(1.9)   \\
128& 20 & 0.285743 & 0.2857428(9) & 14950(4)   \\
   & 20 & 0.285744 & 0.2857438(10) & 14945(4)   \\
192& 10 & 0.285743 & 0.2857430(7) & 33150(12)   \\
   & 10 & 0.285744 & 0.2857435(8) & 33141(13)   \\
\hline
\end{tabular}
\end{center}
\end{table}

\begin{table}
\caption{
MC results for the phenomenological couplings 
at fixed $R_\xi= 0.5943$ for the RSIM at $p=0.8$.
}
\label{tablerun0.8_2}
\footnotesize
\begin{tabular}{@{}rlllllll}
\hline 
\multicolumn{1}{c}{$L$}&
\multicolumn{1}{c}{$\beta_{\rm run}$}&
\multicolumn{1}{c}{$\bar{U}_{22}$}&
\multicolumn{1}{c}{$\bar{U}_4$}&
\multicolumn{1}{c}{$\bar{U}_d$}&
\multicolumn{1}{c}{$\bar{U}_{\rm im}$}&
\multicolumn{1}{c}{$\bar{R'}_\xi$}&
\multicolumn{1}{c}{$-\bar{U'}_4$}\\
\hline 
8  & 0.285744  & 0.1504(6)  & 1.6066(5) & 1.45624(19) & 1.8021(14)    & 18.250(9) & 32.153(20) \\
12 & 0.285744  & 0.1490(4)  & 1.6197(3) & 1.47067(13) & 1.8134(9)     & 32.861(14) & 58.99(3) \\
16 & 0.285744  & 0.1484(5)  & 1.6262(4) & 1.47779(14) & 1.8192(10)    & 49.985(22) & 90.40(5) \\
24 & 0.285744  & 0.1480(6) & 1.6328(5)    & 1.48474(21) & 1.8252(14)  & 90.24(6)   & 164.51(14) \\
32 & 0.285744  & 0.1482(6) & 1.6362(5)   & 1.48798(19) & 1.8289(13) & 137.28(8) & 251.20(20) \\
   & 0.285761  & 0.1495(7) & 1.6370(6)   & 1.48752(22) & 1.8313(15) & 137.12(9) & 251.23(19) \\
48 & 0.285744  & 0.1476(8) & 1.6394(7)  & 1.49179(25) & 1.8313(17)   & 248.29(19) & 456.2(4) \\
   & 0.285748  & 0.1494(6) & 1.6407(5)  & 1.49132(20) & 1.8348(13)   & 248.20(14) & 456.3(3) \\
   & 0.285751  & 0.1497(9) & 1.6409(8)    & 1.49117(27) & 1.8356(20) & 248.04(20) & 456.4(5) \\
64 & 0.285742  & 0.1472(9) & 1.6410(7)   & 1.49380(26) & 1.8324(18) & 378.5(3) & 695.8(8) \\
   & 0.285744  & 0.1475(8) & 1.6412(7)   & 1.49365(26) & 1.8330(17) & 378.7(3) & 696.2(7) \\
96 & 0.285744  & 0.1478(12)& 1.6429(10)& 1.4950(4)   & 1.8351(25) & 684.2(8) & 1259.9(1.9) \\
128& 0.285743  & 0.1477(15) & 1.6440(13)  & 1.4963(5)  & 1.836(3) & 1043.2(1.5) & 1923(4) \\
   & 0.285744  & 0.1471(18) & 1.6434(15) & 1.4963(5)  & 1.835(4) & 1042.7(1.6) & 1921(4) \\
192& 0.285743  & 0.1486(19) & 1.6459(16) & 1.4973(6) & 1.839(4) & 1889(4) & 3493(12) \\
   & 0.285744  & 0.1490(22) & 1.6460(18) & 1.4970(7) & 1.840(5) & 1887(4) & 3485(14) \\
\hline
\end{tabular}
\end{table}

\begin{table}
\caption{
MC results at fixed $R_\xi= 0.5943$ for the RSIM at $p=0.65$.
$N_s$ is the number of samples divided by $1000$.
}
\label{tablerun0.65}
\footnotesize
\begin{tabular}{@{}rrlllllll}
\hline
\multicolumn{1}{c}{$L$}&
\multicolumn{1}{c}{$N_s$}&
\multicolumn{1}{c}{$\beta_{\rm run}$}&
\multicolumn{1}{c}{$\beta_f$}&
\multicolumn{1}{c}{$\chi$}&
\multicolumn{1}{c}{$\bar{U}_{22}$}&
\multicolumn{1}{c}{$\bar{U}_4$}&
\multicolumn{1}{c}{$\bar{U}_{\rm im}$}&
\multicolumn{1}{c}{$\bar{R'}_\xi$}\\
\hline 
8  &  200 & 0.373250 & 0.37372(6) & 50.964(9) & 0.2113(7) & 1.5609(6) & 1.8356(14) & 9.882(13) \\
12 &  200 & 0.371650 & 0.37180(3) & 113.804(16) & 0.2016(6) & 1.5751(5) & 1.8372(12) & 17.581(21) \\
16 &  200 & 0.371050 & 0.371132(19) & 200.92(3) & 0.1956(6) & 1.5827(5) & 1.8369(13) & 26.371(15) \\
   &  200 & 0.371347 & 0.371080(19) & 201.045(23) & 0.1959(6) & 1.5831(5) & 1.8378(13) & 26.360(15) \\
24 &   50 & 0.370500 & 0.370626(21) & 446.94(10) & 0.1896(11) & 1.5926(10) & 1.8390(24) & 46.69(5) \\
   &   50 & 0.370600 & 0.370606(19) & 446.87(13) & 0.1892(12) & 1.5929(10) & 1.839(3) & 46.78(5) \\
   &   50 & 0.370700 & 0.370620(17) & 446.98(9) & 0.1904(11) & 1.5934(10) & 1.8410(24) & 46.66(6) \\
32  & 100 & 0.370420 & 0.370457(9) & 787.67(13) & 0.1856(8) & 1.5984(7) & 1.8397(17) & 71.13(6) \\
    & 100 & 0.370482 & 0.370420(9) & 788.01(12) & 0.1853(9) & 1.5982(7) & 1.8392(18) & 71.11(6) \\
48  &  30 & 0.370290 & 0.370310(8) & 1749.2(4) & 0.1821(14) & 1.6068(11) & 1.844(3) & 124.59(17) \\
    &  30 & 0.370304 & 0.370304(8) & 1750.0(5) & 0.1823(15) & 1.6067(13) & 1.844(3) & 124.39(15) \\
    &  30 & 0.370329 & 0.370304(9) & 1751.2(5) & 0.1822(15) & 1.6061(12) & 1.843(3) & 124.37(15) \\
64  &  30 & 0.370156 & 0.370249(5) & 3079.8(9) & 0.1778(15) & 1.6096(13) & 1.841(3) & 188.1(6) \\
    &  30 & 0.370249 & 0.370243(6) & 3084.6(8) & 0.1761(15) & 1.6081(12) & 1.837(3) & 187.6(6) \\
96  &  20 & 0.370156 & 0.370209(4) & 6839.3(2.1) & 0.1742(15) & 1.6149(13) & 1.841(3) & 335.8(1.0) \\
    &  20 & 0.370209 & 0.370215(3) & 6842.2(2.3) & 0.1734(19) & 1.6142(15) & 1.840(4) & 334.4(1.2) \\
128 &  10 & 0.370196 & 0.370196(3) & 12059(5) & 0.170(3) & 1.6156(22) & 1.837(6) & 504.0(2.2) \\
    &  10 & 0.370209 & 0.370196(3) & 12059(5) & 0.1702(25) & 1.6162(21) & 1.837(5) & 504.8(2.0) \\
\hline
\end{tabular}
\end{table}

\begin{table}
\caption{
MC results at fixed $R_\xi= 0.5943$ for the RBIM at $p=0.55$.
$N_s$ is the number of samples divided by $1000$.
}
\label{tablerun0.55}
\footnotesize
\begin{tabular}{@{}rrlllllll}
\hline
\multicolumn{1}{c}{$L$}&
\multicolumn{1}{c}{$N_s$}&
\multicolumn{1}{c}{$\beta_{\rm run}$}&
\multicolumn{1}{c}{$\beta_f$}&
\multicolumn{1}{c}{$\chi$}&
\multicolumn{1}{c}{$\bar{U}_{22}$}&
\multicolumn{1}{c}{$\bar{U}_4$}&
\multicolumn{1}{c}{$\bar{U}_{\rm im}$}&
\multicolumn{1}{c}{$\bar{R'}_\xi$}\\
\hline 
8 &  640 & 0.432500  & 0.432539(17) & 79.781(4) & 0.12857(22) & 1.60385(18) & 1.7710(5) & 11.270(3) \\
12 & 640 & 0.432340  & 0.432384(10) & 178.251(9) & 0.13629(23) & 1.61874(19) & 1.7959(5) & 20.143(5) \\
16 & 640 & 0.432330  & 0.432335(6) & 314.646(17) & 0.1393(3) & 1.62592(20) & 1.8070(5) & 30.544(7) \\
24 &  256 & 0.432340 & 0.432303(5) & 699.72(6) & 0.1432(4) & 1.6338(3) & 1.8199(8) & 55.036(20) \\
32  & 100 & 0.432300 & 0.432295(5) & 1232.80(18) & 0.1440(7) & 1.6371(6) & 1.8243(14) & 83.76(5) \\
48 &  32 & 0.432293  & 0.432294(5) & 2737.8(7) & 0.1443(12) & 1.6401(10) & 1.828(3) & 151.64(16) \\
64 &  30 & 0.432285  & 0.432290(4) & 4820.2(9)   & 0.1443(10) & 1.6411(8) & 1.8288(21) & 230.27(20) \\
96 &  10 & 0.432279  & 0.432291(4) & 10693(3) & 0.1446(20) & 1.6429(16) & 1.831(4) & 417.9(7) \\
   &  20 & 0.432288  & 0.432290(3) & 10693.6(2.4)& 0.1480(14) & 1.6455(11) & 1.838(3) & 416.6(5) \\
128&  20 & 0.432294  & 0.4322901(17) & 18823(4)  & 0.1472(14) & 1.6457(12) & 1.837(3) & 635.4(7) \\
\hline
\end{tabular}
\end{table}

\begin{table}
\caption{
MC results at fixed $R_\xi= 0.5943$ for the RBIM  at $p=0.7$.
$N_s$ is the number of samples divided by $1000$.
}
\label{tablerun0.7}
\footnotesize
\begin{tabular}{@{}rrllllllll}
\hline
\multicolumn{1}{c}{$L$}&
\multicolumn{1}{c}{$N_s$}&
\multicolumn{1}{c}{$\beta_{\rm run}$}&
\multicolumn{1}{c}{$\beta_f$}&
\multicolumn{1}{c}{$\chi$}&
\multicolumn{1}{c}{$\bar{U}_{22}$}&
\multicolumn{1}{c}{$\bar{U}_4$}&
\multicolumn{1}{c}{$\bar{U}_{\rm im}$}&
\multicolumn{1}{c}{$\bar{R'}_\xi$}\\
\hline 
8  & 640 & 0.325900 & 0.325905(9) & 79.519(3) & 0.08327(15) & 1.61733(12) & 1.7256(3) & 18.308(3) \\
12 & 640 & 0.326200    & 0.326253(5) & 177.657(8) & 0.09027(17) & 1.63479(13) & 1.7521(3) & 33.299(6) \\
16 & 640 & 0.326400    & 0.326409(3) & 313.629(13) & 0.09444(17) & 1.64355(14) & 1.7663(4) & 51.082(8) \\
24 &  250 & 0.326560  & 0.326547(3) & 697.55(5) & 0.0991(3) & 1.65209(24) & 1.7810(6) & 93.642(24) \\
32 & 100 & 0.326600 &  0.326605(3) & 1229.06(14) & 0.1021(5) & 1.6564(4) & 1.7891(10) & 144.08(6) \\
48 &  30 & 0.326660 & 0.326651(3) & 2729.0(5) & 0.1059(9) & 1.6606(7) & 1.7983(18) & 264.59(23) \\
64 &  30 & 0.326629 & 0.3266746(16) & 4802.1(8) & 0.1073(9) & 1.6618(8) & 1.8013(20) & 408.1(8) \\
   &  30 & 0.326640 & 0.3266792(19) & 4802.9(7) & 0.1069(8) & 1.6618(7) & 1.8008(17) & 408.2(6) \\
96 &  20 & 0.326649 & 0.3266919(12) & 10650.6(1.9) & 0.1117(11) & 1.6645(10) & 1.8097(24) & 749.4(2.6) \\
   &  20 & 0.326663 & 0.3266922(13) & 10650.4(2.2) & 0.1122(12) & 1.6655(10) & 1.811(3) & 749.0(2.2) \\
128&  10 & 0.326698 & 0.3266982(11) & 18756(5) & 0.1144(18) & 1.6665(15) & 1.815(4) & 1151(3) \\
   &  10 & 0.326706 & 0.3266977(12) & 18775(5) & 0.1131(17) & 1.6646(14) & 1.812(4) & 1150(3) \\
\hline
\end{tabular}
\end{table}

\section{Finite-size scaling analysis at fixed $R_\xi$ 
of the randomly site-diluted Ising model} 
\label{FSSRSIM}

In this section we describe the FSS analyses of the
MC simulations for the RSIM at $p=0.8$ and $p=0.65$.  The
analyses are performed at a fixed value of $R_\xi$.

\subsection{FSS at fixed phenomenological coupling $R$}
\label{FSSfixedR}

Instead of performing the FSS analysis at fixed Hamiltonian
parameters, we analyze the data at a fixed value $R_f$ of a given
phenomenological coupling $R$, as discussed in Sec.~\ref{sec3.2}.  
The most convenient choice
for the value $R_f$ is $R_f\approx R^*$, where $R^*$ is the asymptotic
value of $R$ at $\beta_c$.
Data at fixed $R=R_f$ are obtained by computing $R(\beta)$ in a
neighborhood of $\beta_c$.  This is done by reweighting the MC data
obtained in a simulation at $\beta = \beta_{\rm run} \approx \beta_c$.
Given $R(\beta)$, one determines the value $\beta_{f}$ such that
$R(\beta = \beta_f) = R_{f}$.  All interesting observables are then
measured at $\beta_f$; their errors at fixed $R=R_f$ are determined by
a standard jackknife analysis.

In the following we present the FSS analysis at fixed $R_\xi\equiv\xi/L$,
choosing $R_{\xi,f}=0.5943$, which is the estimate of $R_\xi^*$
obtained in Ref.~\cite{CMPV-03}: $R_\xi^*=0.5943(9)$.  In
Tables~\ref{tablerun0.8_1}, \ref{tablerun0.8_2}, 
and \ref{tablerun0.65} we report the results
obtained for the RSIM at $p=0.8$ and $p=0.65$, respectively.  We also
performed FSS analyses at fixed $U_4=U_{4,f}=1.650$ [from
Ref.~\cite{CMPV-03} that quotes $U_4^*=1.650(9)$]. 
We do not report the corresponding
results because they are consistent with, though slightly less precise than,
those obtained at fixed $R_\xi=0.5943$.

FSS at fixed $R$ has the advantage that it does not require a precise
knowledge of the critical value $\beta_c$. But there is another nice side
effect: for some observables the statistical errors at fixed $R_f$ are smaller
than those at fixed $\beta$ (close to $\beta_c$).  
For example, in the case of the RSIM at $p=0.8$, 
we find
\begin{eqnarray}
{\mbox{err}[\chi|_{\beta_c}] \over \mbox{err}[\chi|_{R_\xi=0.5943}] }
    \approx 10,    &\qquad&
{\mbox{err}[\chi|_{\beta_c}] \over \mbox{err}[\chi|_{U_4=1.650}]}
    \approx 1.7,    \\
{\mbox{err}[U_4|_{\beta_c}]\over \mbox{err}[U_4|_{R_\xi=0.5943}]}
    \approx 1.7,    &\qquad&
{\mbox{err}[U_{22}|_{\beta_c}]\over\mbox{err}[U_{22}|_{R_\xi=0.5943}]}
    \approx 1.1,    \nonumber \\
{\mbox{err}[\chi'|_{\beta_c}]\over \mbox{err}[\chi'|_{R_\xi=0.5943}]}
    \approx 2.0,    &\qquad&
{\mbox{err}[R'_{\xi}|_{\beta_c}]\over\mbox{err}[R'_\xi|_{R_\xi=0.5943}]}
    \approx 1.8,    \nonumber
\end{eqnarray}
which are approximately independent of $L$.
Similar numbers are found for the other models that we have simulated.

\subsection{Universal values of $U_{22}$ and $U_4$ at fixed $R_\xi=0.5943$}
\label{u22u4}

\begin{figure}[tb]
\vspace{1cm}
\centerline{\psfig{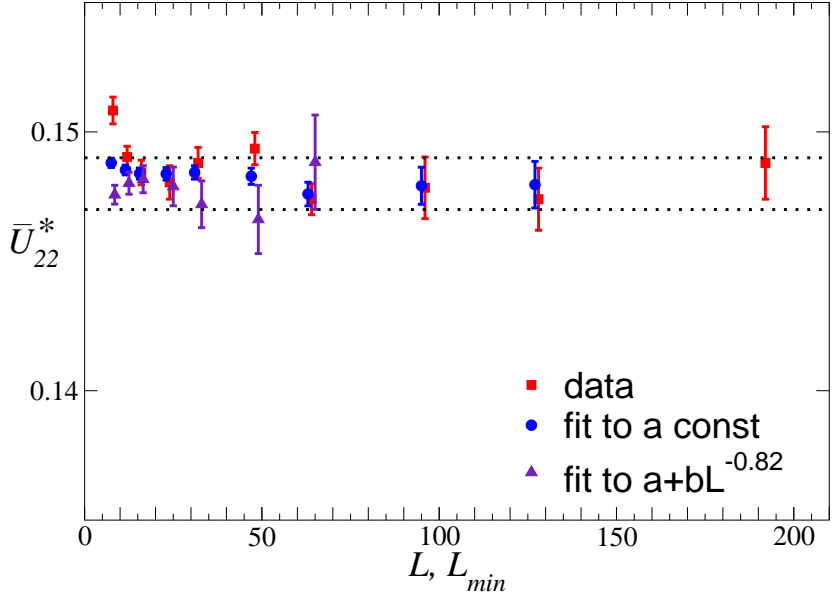}}
\vspace{2mm}
\caption{
MC results $\bar{U}_{22}(L)$ and estimates of
$\bar{U}_{22}^*(L_{\rm min})$ as obtained in different fits at fixed
$R_\xi=0.5943$ for the RSIM at $p=0.8$.
Some data are slightly shifted along the $x$-axis to make them visible.
On the $x$ axis we report $L$ when plotting the MC data and 
the minimum lattice size $L_{\rm min}$ used in the fit
when plotting the fit results.
The dotted lines correspond to the final estimate $\bar{U}_{22}^*=0.148(1)$.
}
\label{u22barp0.8}
\end{figure}

\begin{figure}[tb]
\vspace{1cm}
\centerline{\psfig{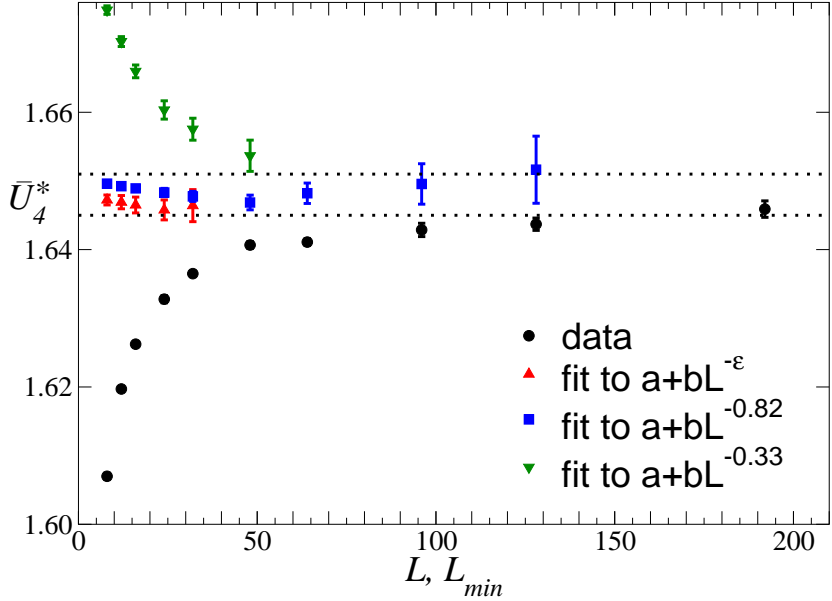}}
\vspace{2mm}
\caption{
MC results $\bar{U}_{4}(L)$ and estimates of
$\bar{U}_{4}^*(L_{\rm min})$ as obtained in different fits at fixed
$R_\xi=0.5943$ for the RSIM at $p=0.8$.
In the second fit  
($a + b L^{-\epsilon}$) $\epsilon$ is a free parameter.
As in Fig.~\protect\ref{u22barp0.8}, the $x$ axis corresponds to $L$
(MC data) and $L_{\rm min}$ (fit results).
The dotted lines correspond to the final estimate $\bar{U}_{4}^*=1.648(3)$.
}
\label{u4barp0.8}
\end{figure}

\begin{figure}[tb]
\vspace{1cm}
\centerline{\psfig{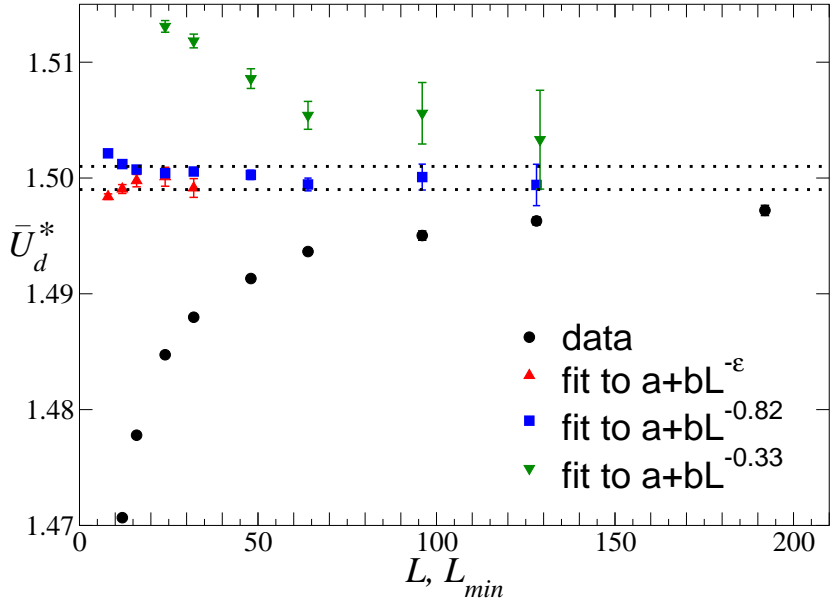}}
\vspace{2mm}
\caption{
MC results $\bar{U}_{d}(L)$ and estimates of
$\bar{U}_{d}^*(L_{\rm min})$ as obtained in different fits at fixed
$R_\xi=0.5943$ for the RSIM at $p=0.8$. 
In the second fit  
($a + b L^{-\epsilon}$) $\epsilon$ is a free parameter.
As in Fig.~\protect\ref{u22barp0.8}, the $x$ axis corresponds to $L$
(MC data) and $L_{\rm min}$ (fit results).
The dotted lines correspond to the final estimate $\bar{U}_d^*=1.500(1)$.
}
\label{udbarp0.8}
\end{figure}

Here we determine the universal large-$L$ limits 
of the quartic cumulants $U_{22}$ and $U_4$ at fixed $R_\xi=0.5943$---we
call them $\bar{U}_{22}$ and $\bar{U}_{4}$, respectively---using 
the data at $p=0.8$ reported in Table~\ref{tablerun0.8_2}.
The data of $\bar{U}_{22}$ in the range $8\le L \le 192$ shown in
Fig.~\ref{u22barp0.8} have a very small $L$ dependence.
In particular, there is no
evidence of scaling corrections associated with the leading
correction-to-scaling exponent $\omega \approx 0.3$.  This confirms earlier
results \cite{BFMMPR-98,Hukushima-00,CMPV-03} indicating that leading scaling
corrections in the RSIM at $p=0.8$ are very small. Moreover,
also the next-to-leading scaling corrections associated with
$\omega_2\approx 0.8$ are quite small. Indeed, if we fit the data with 
$L\ge L_{\rm min}$ to a constant, a fit with $\chi^2/{\rm DOF}\lesssim 1$ (DOF is the 
number of degrees of freedom of the fit) is already
obtained for $L_{\rm min} = 12$. The corresponding estimates of 
$\bar{U}_{22}^*$ are independent of $L_{\rm min}$ within error bars,
see Fig.~\ref{u22barp0.8}. We also fit the data to 
$a+bL^{-\epsilon}$ with $\epsilon=0.33$ and $\epsilon=0.82$,
which are the best MC estimate of $\omega$ (see Sec.~\ref{estomega}) and 
the FT estimate of $\omega_2$ (see \ref{omega2}), respectively. In both cases 
$b\approx 0$ within errors already for $L_{\rm min} = 16$. 
Finally, we also do fits keeping $\epsilon$ as a free parameter. 
The estimates of $\bar{U}_{22}^*$ are independent of $L_{\rm min}$ 
within error bars; for $L_{\rm min} = 8$ we get $\bar{U}_{22}^* = 0.1483(3)$,
with $\chi^2/{\rm DOF} = 0.8$. The 
estimates of $\bar{U}_{22}^*$ obtained in the fits with $\epsilon=0.82$
are also plotted in Fig.~\ref{u22barp0.8} versus the minimum
lattice size $L_{\rm min}$ of the data considered in the fits.
Collecting results, we obtain the final estimate
\begin{equation}
\bar{U}_{22}^* = 0.148(1).
\label{u22barest}
\end{equation}
We perform a similar analysis of $\bar{U}_4$.  The data are shown in 
Fig.~\ref{u4barp0.8}.  In this case, there is clear evidence of scaling
corrections.
A fit of the data to $a+bL^{-\epsilon}$, taking  $\epsilon$ as a free parameter,
gives $\bar{U}_{4}^*=1.6472(7)$ and $\epsilon = 0.95(5)$  for $L_{\rm min} = 8$: 
There is no evidence of scaling
corrections with exponent $\omega\approx 0.3$. This is confirmed by 
fits to $a+bL^{-\epsilon}$ with $\epsilon = 0.3$ fixed. The leading term $a$ varies
significantly with $L_{\rm min}$, exactly as the original data
(see Fig.~\ref{u4barp0.8}).  The numerical results for $\bar{U}_4$ are 
much better described by assuming that the leading scaling corrections
are proportional to $L^{-\omega_2} \sim L^{-0.82}$. A fit 
to $a+bL^{-0.82}$ gives estimates of $\bar{U}_4^*$ that show a very tiny 
dependence on $L_{\rm min}$, see Fig.~\ref{u4barp0.8}. The dependence on
$\omega_2$ is also small: if $\omega_2$ varies between
$0.74$ and $0.90$ [this corresponds to considering the FT estimate
$\omega_2 = 0.82(8)$], estimates change by less than 0.001 for 
$L_{\rm min} \le 32$.  These analyses lead to the final estimate
\begin{equation}
\bar{U}_{4}^* = 1.648(3).
\label{u4barest}
\end{equation}
We finally consider the
difference $\bar{U}_d\equiv \bar{U}_4-\bar{U}_{22}$, whose data are
very precise because of an unexpected cancellation of the statistical
fluctuations, see Table~\ref{tablerun0.8_2}.  
In Fig.~\ref{udbarp0.8} we show the results of the same
fits as done for $\bar{U}_4$. They lead to the estimate
\begin{equation}
\bar{U}_{d}^* = 1.500(1),
\label{udbarest}
\end{equation}
which is perfectly consistent with the estimates obtained for
$\bar{U}_{22}^*$ and $\bar{U}_{4}^*$.

\subsection{Estimate of the leading correction-to-scaling exponent $\omega$}
\label{estomega}

\begin{figure}[tb]
\vspace{1cm}
\centerline{\psfig{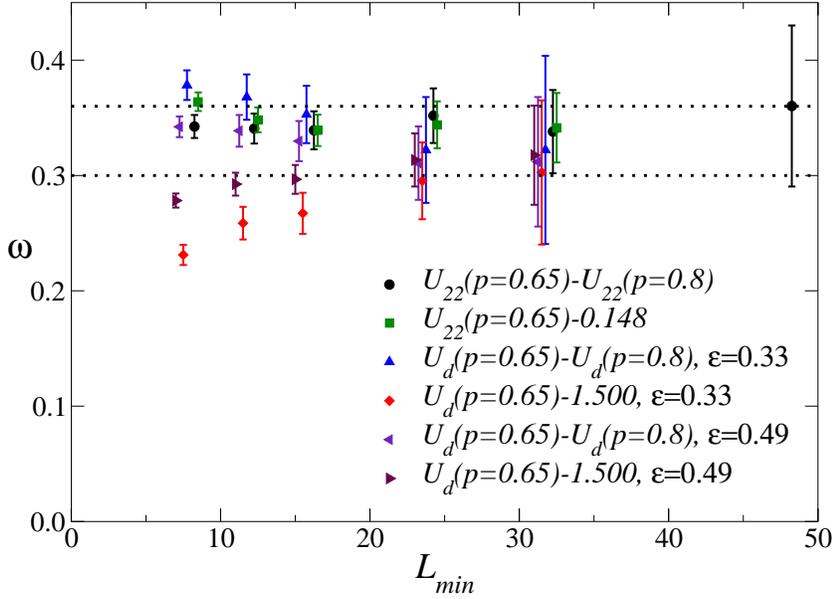}}
\vspace{2mm}
\caption{
Estimates of $\omega$ from fits of 
$\Delta_{22,a} \equiv \bar{U}_{22}(p=0.65) - \bar{U}_{22}(p=0.8)$,
$\Delta_{22,b} \equiv \bar{U}_{22}(p=0.65) - \bar{U}_{22}^*$,
$\Delta_{d,a} \equiv  \bar{U}_{d}(p=0.65) - \bar{U}_{d}(p=0.8)$,
$\Delta_{d,b} \equiv \bar{U}_{d}(p=0.65) - \bar{U}_d^*$,
as a function of $L_{\rm min}$, which is the minimum
lattice size considered in the fits. $\Delta_{22,a}$ and 
$\Delta_{22,b}$ are fitted as $\ln \Delta = a - \omega \ln L$; 
$\Delta_{d,a}$ and
$\Delta_{d,b}$ are fitted as $\ln \Delta = a - \omega \ln L + b L^{-\epsilon}$.
Some data are slightly shifted along the $x$-axis to make them visible.
The dotted lines correspond to the final estimate $\omega=0.33(3)$.
}
\label{omega1}
\end{figure}

In order to estimate $\omega$, we use the data at $p=0.65$. We
consider the differences
\begin{eqnarray}
&&\Delta_{22,a} \equiv \bar{U}_{22}(p=0.65;L) - \bar{U}_{22}(p=0.8;L) ,
\label{deltau22d}\\
&&\Delta_{d,a} \equiv \bar{U}_{d}(p=0.65;L) - \bar{U}_{d}(p=0.8;L), 
\end{eqnarray}
and also
\begin{eqnarray}
&&\Delta_{22,b} \equiv \bar{U}_{22}(p=0.65;L) - \bar{U}_{22}^*,
\label{deltau22db}\\
&&\Delta_{d,b} \equiv \bar{U}_{d}(p=0.65;L) - \bar{U}_{d}^*,
\end{eqnarray}
where $\bar{U}_{22}^*$ and $\bar{U}_{d}^*$ are given in Eqs.~(\ref{u22barest}) and
(\ref{udbarest}).
Because of the universality 
of the large-$L$ limit of the phenomenological 
couplings---hence, they do not depend on $p$---they behave as
\begin{equation}
\Delta\approx 
c_{\Delta,11} L^{-\omega} + c_{\Delta,12} L^{-2\omega} + \cdots + 
  c_{\Delta,21} L^{-\omega_2} +\cdots
\label{eq-sec5.3}
\end{equation}
The quantities defined in Eqs.~(\ref{deltau22d}) and 
(\ref{deltau22db}) are well fitted to $c_{\Delta,11}
L^{-\omega}$.\footnote{We will often say that we fit a quantity 
${\cal O}$ to $a L^x$ or to $aL^x(1 + b L^y)$, $y < 0$. 
What we really do is a fit of $\ln {\cal O}$ to 
$\ln a + x \ln L$ and to 
$\ln a + x \ln L + b L^{y}$.}
For instance, the analysis of $\Delta_{22,a}$ gives
$\omega = 0.342(10)$ and $\omega = 0.352(24)$ for 
$L_{\rm min} = 8$ and 24, respectively; in both cases $\chi^2/{\rm DOF} \approx 0.5$.
Such a fit instead gives a relatively large 
$\chi^2/{\rm DOF}$ ($\chi^2/{\rm DOF}=2.3$  for $L_{\rm min} = 12$)
when applied to $\bar{U}_{d}$, essentially
because the data of $\bar{U}_{d}$ have a better relative precision.
Moreover, a clear systematic drift is observed when varying $L_{\rm min}$.
Therefore, we must include the next-to-leading corrections. 
We thus performed fits of the form $c_{\Delta,11}
L^{-\omega} (1 + d L^{-\epsilon}$), where $\epsilon$ is an effective 
exponents that takes into account two next-to-leading corrections 
and should vary in $[\omega,\omega_2-\omega]$. Given the results obtained 
from the analysis of $\bar{U}_{22}$ and the FT estimate $\omega_2 = 0.82(8)$,
we have taken $\epsilon \in [0.3,0.6]$.
The dependence on $\epsilon$ is small: for instance, 
for $L_{\rm min} = 16$, the analysis of $\Delta_{d,a}$ gives 
$\omega = 0.35(3)$, 0.33(2), 0.32(2) for $\epsilon = 0.3$, 0.5, 0.6.
The results corresponding
to $\epsilon = 0.33$ and $\epsilon = 0.49 \approx \omega_2 - \omega$ 
are reported in Fig.~\ref{omega1} as a function of 
$L_{\rm min}$, the smallest lattice size used in the analysis.
They lead to the estimate
\begin{equation}
\omega=0.33(3),
\end{equation} 
which is in agreement with the FT six-loop result
\cite{PV-00} $\omega=0.25(10)$ (we also mention the five-loop result
$\omega = 0.32(6)$ of Ref.~\cite{PS-00})
and with  the MC result \cite{BFMMPR-98} $\omega=0.37(6)$.

\begin{figure}[tb]
\vspace{1cm}
\centerline{\psfig{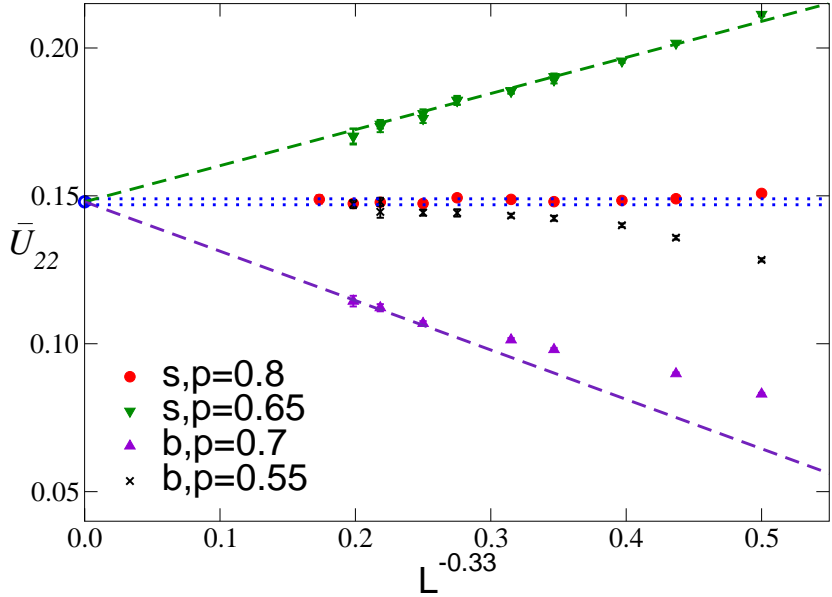}}
\vspace{2mm} \caption{ 
Plot of $\bar{U}_{22}$ versus $L^{-\omega}$ with
$\omega=0.33$ for the RSIM (s) at $p=0.8$ and $p=0.65$
and for the RBIM (b) at $p=0.7$ and $p=0.55$.
}
\label{u22omega}
\end{figure}

In writing Eq.~(\ref{eq-sec5.3}) 
we assumed that the scaling limit does not 
depend on $p$. We can now perform a consistenty check, verifying whether 
$\bar{U}_{22}$ and $\bar{U}_4$  for $p=0.65$ converge to the estimates 
(\ref{u22barest}) and (\ref{u4barest}). A fit of $\bar{U}_{22}$ to 
$\bar{U}_{22}^* + c_{22,11} L^{-\omega} + c_{22,2} L^{-\epsilon}$ 
gives $\bar{U}_{22}^* = 0.154(4)$, 0.152(2), 0.149(6), 0.148(5) for
$(L_{\rm min},\epsilon) = (8,2\omega)$, $(8,\omega_2)$,
$(12,2\omega)$, $(12,\omega_2)$, respectively. Here we used $\omega = 0.33$
as determined above and $\omega_2 = 0.82$. These results are in  
agreement with the estimate  (\ref{u22barest}). Such an agreement is 
also clear from Fig.~\ref{u22omega}, where 
we plot $\bar{U}_{22}$ versus $L^{-\omega}$.
The same analysis 
applied to $\bar{U}_4$ gives $\bar{U}_{4}^* = 1.640(4)$, 
1.644(3), 1.640(5), 1.644(4), for the same values of $(L_{\rm min},\epsilon)$, 
which are compatible with the estimate  (\ref{u4barest}).

\subsection{Determination of the improved RSIM}
\label{pstar}

We now estimate the value $p^*$ of the spin concentration that corresponds to 
an improved model: for $p=p^*$ the leading scaling corrections with exponent 
$\omega$ vanish. We already know that the RSIM with $p=0.8$ is approximately
improved, so that $p^* \approx 0.8$.  In the following we make
this statement more precise. 
We consider $\bar{U}_{22}$, which 
has small next-to-leading scaling corrections, and 
determine the value of $p$ at
which the leading scaling corrections to this quantity vanish. 
As we remarked in Sec.~\ref{FSS-general},
the same cancellation occurs in any other quantity.

\begin{figure}[tb]
\vspace{1cm}
\centerline{\psfig{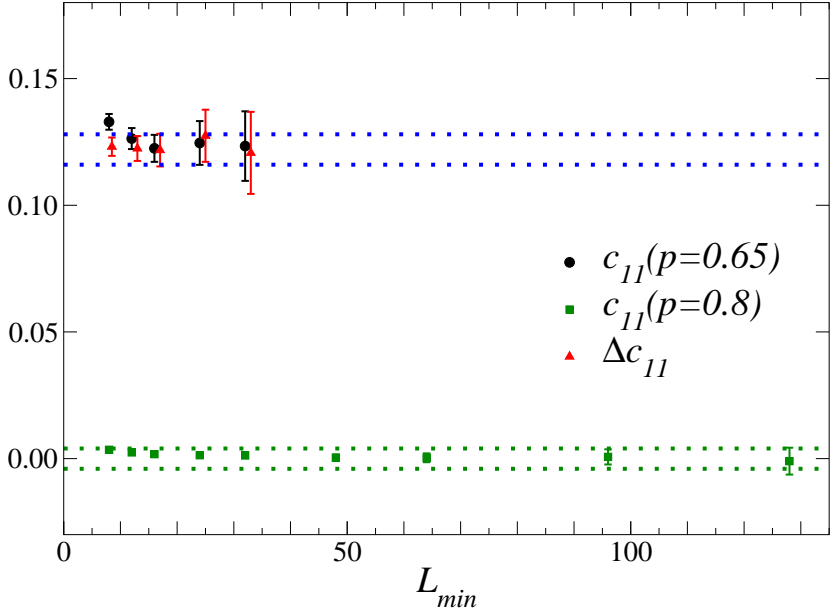}}
\vspace{2mm}
\caption{
Results of the fits to estimate the amplitude of the leading
scaling correction in $\bar{U}_{22}$.
}
\label{c11fig}
\end{figure}

To determine $p^*$, 
we fit the data of $\bar{U}_{22}$ at $p=0.8$ and $p=0.65$ to
$\bar{U}_{22}^* + c_{22,11}(p) L^{-\epsilon}$, and the difference
$\Delta_{22,a}$ defined in Eq.~(\ref{deltau22d}) to 
$c_{\Delta,11} L^{-\epsilon}$. We have performed fits in which 
$\epsilon$ is fixed, $\epsilon=\omega = 0.33(3)$, and fits in which it is taken as 
a free parameter. 
The results of the fits with $\epsilon=\omega$ are shown in Fig.~\ref{c11fig}
for several values of $L_{\rm min}$, the smallest lattice size included in
the fit.  We obtain 
\begin{eqnarray}
&&c_{\Delta,11} = 0.122(6), \nonumber \\ 
&&c_{22,11}(p=0.65) = 0.122(6), \label{ccorr} \\
&&c_{22,11}(p=0.8) = 0.000(4), \nonumber 
\end{eqnarray}
where the first two estimates correspond to $L_{\rm min} = 16$ and the last 
one to $L_{\rm min} = 48$;
errors are such to include the results of all fits.
These results give us the upper bound 
\begin{equation}
  \left| {c_{22,11}(p=0.8) \over c_{22,11}(p=0.65)}\right|
  \lesssim {4\over122} \approx {1\over 30}.
\label{bound-c11-1}
\end{equation}
For $p=0.8$ scaling corrections are at least a factor of 30 smaller than
those occurring for $p = 0.65$. This bound 
will be useful in the following to assess the relevance of the 
``systematic error" in the fits
of the data at $p=0.8$ due to possible residual leading scaling corrections.
Indeed, the ratio that appears in Eq.~(\ref{bound-c11-1}) 
does not depend on the 
observable. In the notations of Sec.~\ref{FSS}, 
$c_{22,11}(p) = r_{3,0}(z_f) u_3(t=0,p)$ where $r_{3,0}(z_f)$ is a model-independent 
constant that depends on the observable (in this case on $\bar{U}_{22}$). 
The constant $r_{3,0}(z_f)$ drops out in the ratio, since
\begin{equation}
{c_{22,11}(p_1)\over c_{22,11}(p_2)} = {u_3(t=0,p_1) \over u_3(t=0,p_2)}.
\end{equation}
Therefore, 
given a generic observable ${\cal O}(p)$ that behaves as 
\begin{equation} 
{\cal O}(p) = a(p) L^\sigma (1 + c_{{\cal O} ,11}(p) L^{-\omega} + \ldots) ,
\end{equation}
we have in all cases 
\begin{equation}
 |c_{{\cal O} ,11}(p=0.8)/c_{{\cal O} ,11}(p=0.65)| 
\lesssim 1/30. 
\label{bound-c11}
\end{equation}
Estimates (\ref{ccorr}) allow us to estimate $p^*$. 
We obtain $p^* \approx 0.80$. Since, 
$c_{22,11}(p) = a (p-p^*)$ close to $p\approx 0.80$, the error 
on $p^*$ is simply ${\rm err}[c_{22,11}(p=0.8)]/|a|$. The constant $a$---we 
only need a rough estimate since it is only relevant for the error on 
$p^*$---is determined as 
\begin{equation}
a = \left. {dc_{22,11}\over dp} \right|_{p=p^*}
    \approx {c_{22,11}(p=0.8) - c_{22,11}(p=0.65)  \over 
           0.8 - 0.65} = -0.81(4),
\label{approxdcdp}
\end{equation}
where the reported error is obtained from those of 
$c_{22,11}(p=0.8)$ and $c_{22,11}(p=0.65)$.
This gives
\begin{equation}
p^*=0.800(5) .
\end{equation}
We are not able to assess the error on $a$ due to the
approximation (\ref{approxdcdp}). Note, however, that the dependence on $a$ is small.
If we vary $a$ by a factor of 2, $p^*$ changes at most by 0.01.

\subsection{Determination of the critical temperatures}
\label{betacRSIM}

We determine the critical temperature by
extrapolating the estimates of $\beta_f$ reported in
Tables~\ref{tablerun0.8_1} and \ref{tablerun0.65}. According
to the discussion reported in Sec.~\ref{FSS}, since we have chosen
$R_\xi=0.5943\approx R_\xi^*$,
we expect in general that $\beta_f-\beta_c=O(L^{-1/\nu-\omega})$.
For $p=0.8$, since the model is improved, the leading scaling
corrections are related to the next-to-leading exponent $\omega_2$.
Thus, in this case $\beta_f-\beta_c=O(L^{-1/\nu-\omega_2})$

\begin{figure}[tb]
\vspace{1cm}
\centerline{\psfig{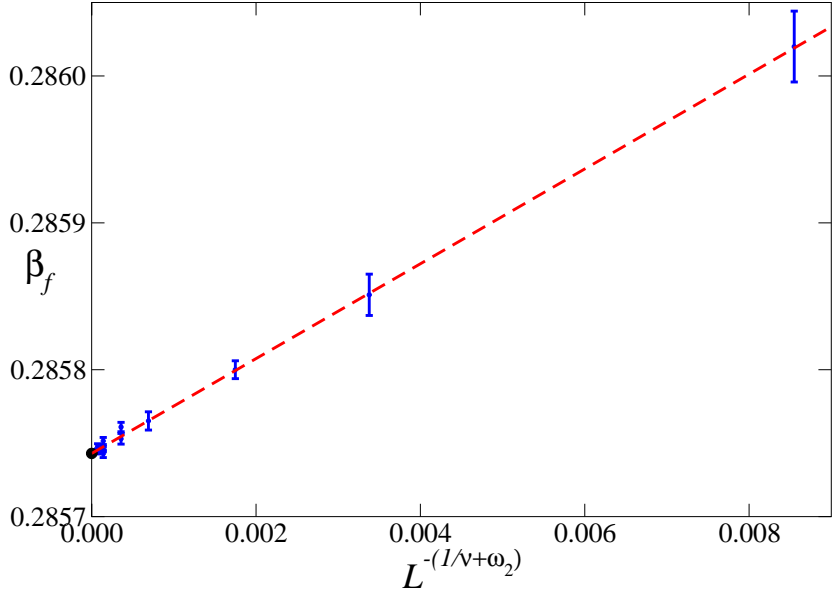}}
\vspace{2mm}
\caption{
Estimates of $\beta_f$ for $p=0.8$ versus $L^{-(1/\nu+\omega_2)}$,
for $\omega_2 = 0.82$ and $\nu = 0.68$.
The dashed line corresponds to a linear fit of the data.
}
\label{bffig}
\end{figure}

In Fig.~\ref{bffig} we show the data for $p=0.8$ versus
$L^{-1/\nu-\omega_2}$ taking $\nu=0.68$ and $\omega_2=0.82$.  The expected
linear behavior is clearly observed. A fit to $\beta_c+ cL^{-1/\nu-\omega_2}$
gives $\beta_c=0.2857429(4)$. We have also taken into account
the error on $\omega_2$ and on $\nu$ (the error due to the 
variation of $\nu$ is essentially negligible compared to the 
first one). This result improves the estimate of $\beta_c$
obtained in Ref.~\cite{CMPV-03}, i.e. $\beta_c=0.285744(2)$.

For $p=0.65$ we fit $\beta_f$ to $\beta_c+ cL^{-1/\nu-\omega} + d
L^{-1/\nu-\epsilon}$ with $\nu = 0.68$, $\omega= 0.33(3)$, $\epsilon\in
[0.6,0.9]$. We obtain $\beta_c=0.370174(3)$. This is consistent with the
estimate $\beta_c=0.370166(6)$ ($L_{\rm min} = 16$) reported in
Ref.~\cite{BFMMPR-98}.

\subsection{Improved phenomenological couplings}
\label{uo}

Beside considering improved Hamiltonians---in these particular models
any thermodynamic
quantity does not have leading scaling corrections---one may also
consider improved observables which are such that the leading scaling 
correction vanishes 
for any Hamiltonian.  Here we determine an improved phenomenological coupling 
by taking an appropriate combination of the cumulants $\bar{U}_4$ and
$\bar{U}_{22}$, i.e. we consider
\begin{equation}
\bar{U}_{\rm im} \equiv  \bar{U}_4 + c_{\rm im} \bar{U}_{22}.
\label{uim}
\end{equation}
In the scaling limit we have generically 
$\bar{U}_\# \approx \bar{U}_\#^* + c_{\#,11} L^{-\omega}$. The constant 
$c_{\rm im}$ is determined by requiring $c_{{\rm im},11} = 0$, 
which gives
\begin{equation}
c_{\rm im} = - {c_{4,11}(p)\over c_{22,11}(p)},
\label{cim}
\end{equation}
where we have written explicitly the $p$ dependence of the coefficients
$c_{4,11}(p)$ and $c_{22,11}(p)$. As discussed in Sec.~\ref{FSS-general},
the ratio (\ref{cim}) is universal within the RDIs universality
class and, in particular, independent of $p$.  
Indeed, $c_{4,11} = r_{U_4,3,0}(z_f) u_3$, $c_{22,11} = r_{U_{22},3,0}(z_f) u_3$,
where $r_{U_4,3,0}(z)$ and $r_{U_{22},3,0}(z)$ are the universal scaling 
functions defined in Sec.~\ref{FSS-general}.
The model-dependent scaling field 
$u_3$ cancels out in the ratio, proving its universality.
The combination
$\bar{U}_{\rm im}$ with the choice (\ref{cim}) has no leading
scaling corrections associated with the exponent $\omega$, so that, 
see Sec.~\ref{FSS-general}, $\bar{U}_{\rm im} = \bar{U}_{\rm im}^* + 
O(L^{-2\omega},L^{-\omega_2})$.

\begin{figure}[tb]
\vspace{1cm}
\centerline{\psfig{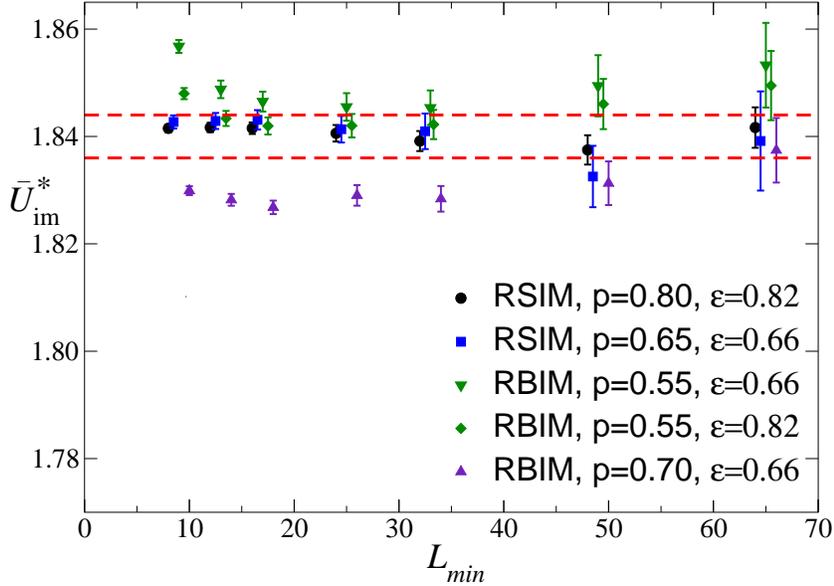}}
\vspace{2mm}
\caption{
Estimates of $\bar{U}_{\rm im}^*$ obtained from fits of $\bar{U}_{\rm im}$ to
$\bar{U}_{\rm im}^*+ b L^{-\epsilon}$.
The dashed lines correspond to the estimate $\bar{U}_{\rm im}^*=1.840(4)$ 
obtained from fits of $\bar{U}_4$, $\bar{U}_{22}$, $\bar{U}_d$ at 
$p = 0.8$.   
}
\label{uimfits}
\end{figure}

The ratio $c_{\rm im}$ can be estimated by determining the leading
scaling correction amplitudes of $\bar{U}_{22}$ and $\bar{U}_4$.
Alternatively, one may estimate it from the ratio
\begin{equation}
{ \bar{U}_{4}(p=0.65;L) - \bar{U}_{4}(p=0.8;L) \over
\bar{U}_{22}(p=0.65;L) - \bar{U}_{22}(p=0.8;L) }
= - c_{\rm im} + a L^{-\omega} + b L^{-\omega_2+\omega} + \cdots
\end{equation}
Both analyses give consistent results, leading to the estimate
\begin{equation}
c_{\rm im} = 1.3(1).
\end{equation}
Therefore,  
\begin{equation}
\bar{U}_{\rm im} = \bar{U}_{4}  + 1.3 \bar{U}_{22}
\label{uimdef}
\end{equation}
has (approximately) vanishing leading scaling corrections for any model and 
any $p$.  More precisely, we have the upper bound $|c_{{\rm im},11}| \lesssim 
0.1|c_{22,11}|$. Since  
$c_{4,11} = - c_{\rm im} c_{22,11}$ and 
$c_{d,11} = - (1 + c_{\rm im}) c_{22,11}$,
we also have 
$|c_{{\rm im},11}|\lesssim 0.1|c_{4,11}|$ and
$|c_{{\rm im},11}|\lesssim 0.05|c_{d,11}|$.
The leading scaling corrections in $\bar{U}_{\rm im}$ are at least a factor of 10 
smaller than those occurring in $\bar{U}_{22}$ and $\bar{U}_{4}$ and 
a factor of 20 smaller than those occurring in $\bar{U}_d$.
This is confirmed by a direct analysis of the data of
$\Delta_{{\rm im},b}$, defined as in Eq.~(\ref{deltau22db}), at $p = 0.65$.
A fit to $c_{{\rm im},11} L^{-\omega}$, $\omega = 0.33$,
gives $|c_{{\rm im},11}|\lesssim 0.005$, to be compared with 
$c_{22,11} = 0.122(6)$ obtained in Sec.~\ref{estomega}. 
In Fig.~\ref{uimfits} we show
results of fits of $\bar{U}_{\rm im}$ to $\bar{U}_{\rm
im}^*+ c L^{-\epsilon}$ with $\epsilon=0.66\approx 2\omega$ and 
$\epsilon=0.82\approx \omega_2$. They
are consistent with the estimate $\bar{U}_{\rm im}^*=1.840(4)$, which
can be obtained from the estimates of $\bar{U}_{4}$, $\bar{U}_{22}$,
and $\bar{U}_{d}$ obtained in Sec.~\ref{u22u4}.  The improved quantity
$\bar{U}_{\rm im}$ is particularly useful to check universality,
because it is less affected by scaling corrections.

We should note that $\bar{U}_{\rm im}$ is useful in generic models in which 
$L^{-\omega}$ corrections are generically present, but is not the optimal quantity
in improved models in which the leading scaling corrections are proportional to 
$c_{21} L^{-\omega_2}$. The coefficients $c_{21}$ can be estimated from the 
data at $p = 0.8$, obtaining
\begin{eqnarray}
&& c_{22,21} = 0.01 (2), \nonumber \\
&& c_{4,21}  = -0.21 (5), \nonumber \\
&& c_{d,21}  = -0.22 (5), \nonumber \\
&& c_{{\rm im},21}  = -0.20 (5) .
\end{eqnarray}
Since the ratios $c_{a,21}/c_{b,21} = r_{a,4,0}(z_f)/r_{b,4,0}(z_f)$ are universal,
the coefficient $c_{22,21}$ is at least a factor of 5 smaller than 
the corresponding one in $\bar{U}_{\rm im}$, $\bar{U}_{4}$, $\bar{U}_{d}$ 
in any RDIs model.
Hence, in improved models $\bar{U}_{22}$ and not $\bar{U}_{\rm im}$ is the 
optimal quantity.

\subsection{The critical exponent $\nu$}
\label{nusec}

We estimate the critical exponent $\nu$ by using the data at $p=0.8$, since in
this case there are no leading corrections to scaling---the model is
improved---the data have smaller errors, and we have results for larger
lattices. We analyze the derivatives $R_\xi'$ and $U_4'$ at fixed $R_\xi$.

\begin{figure}[tb]
\vspace{1cm}
\centerline{\psfig{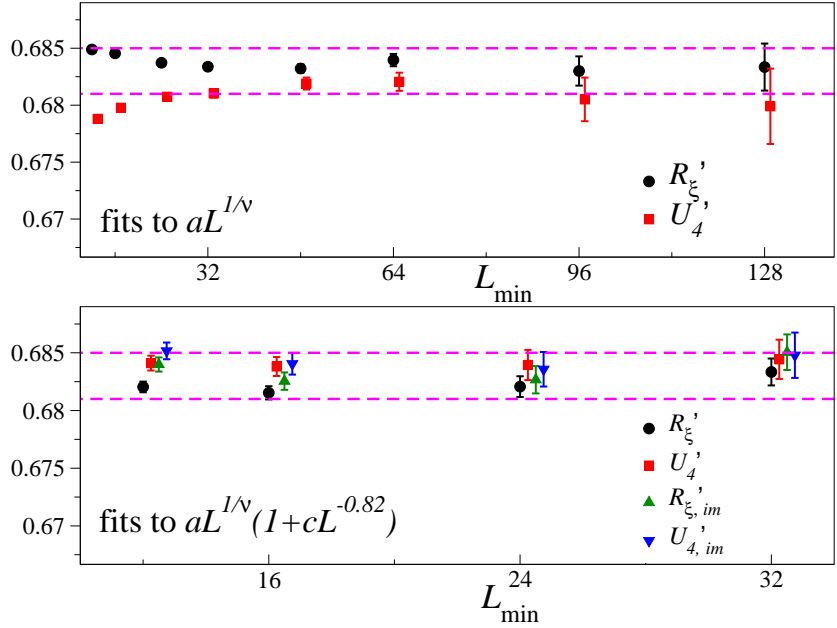}}
\vspace{2mm}
\caption{
  Estimates of the critical exponent $\nu$, obtained by fitting the MC data
  for the RSIM at $p=0.8$ to a simple power law (above), i.e. to $aL^{1/\nu}$,
  and to Eq.~(\ref{fitans}) with $\omega_2=0.82$.  We consider
  $R_\xi'$, $U_4'$, $R'_{\xi,{\rm im}}$ and $U'_{4,{\rm im}}$.  
  Some results are slightly shifted along the $x$-axis to
  make them visible. The dotted lines correspond to the final estimate
  $\nu=0.683(2)$.  }
\label{nuest}
\end{figure}

Since we have established that the RSIM at $p=0.8$ is improved, the dominant
scaling corrections are associated with the next-to-leading exponent
$\omega_2$.  Therefore, we fit $R'$ to $a L^{1/\nu}$ and to
\begin{equation}
a L^{1/\nu} \left ( 1 + c L^{-\omega_2} \right)
\label{fitans}
\end{equation}
or, more precisely, $\ln R'$ to $\ln a + {1\over \nu} \ln L$ or to 
$\ln a + {1\over \nu} \ln L + c 
L^{-\omega_2}$. The exponent $\omega_2$ is fixed to the FT value $\omega_2 =
0.82(8)$.  The results are reported in Fig.~\ref{nuest} as a function of
$L_{\rm min}$.  These fits are quite good, with $\chi^2/{\rm DOF}\lesssim 1$
already for relatively smal values of $L_{\rm min}$. 
The dependence on $\omega_2$ is negligible when $\omega_2$ varies
within the range allowed by the FT estimate.
The fits of $R'_\xi$ to a simple power law are quite stable. The
results for $L_{\rm min}\ge 48$ provide the estimate
\begin{equation}
\nu=0.6835(10).   
\label{nuome2}
\end{equation}

The exponent $\nu$ can also be determined by analyzing the ratio ${\chi'/
  \chi}$, where $\chi'$ is the derivative of $\chi$ with respect to $\beta$.
This ratio also has the asymptotic behavior~(\ref{fitans}).  The corresponding
results are in perfect agreement with those obtained by using $R'_\xi$ and $U'_4$.

Collecting results we obtain the final estimate
\begin{equation}
\nu=0.683(2),
\label{nuestimate}
\end{equation}
which takes into account the fits of  $R'_{\xi}$ and $U'_4$, with and without
the scaling correction with exponent $\omega_2$.
In the determination of the estimate (\ref{nuestimate}), 
we have implicitly assumed 
that the RSIM at $p = 0.8$ is exactly improved so that there are no leading scaling 
corrections. However, $p^*$ is only known approximately and thus some residual
leading scaling corrections may still be present. 
To determine their relevance, we use the 
upper bound (\ref{bound-c11}) and the MC results for $R'$ at $p=0.65$. If the 
leading scaling corrections do not vanish, $R'$ behaves as 
\begin{equation}
R' = a L^{1/\nu}\left(
  1 + b_{R',11} L^{-\omega} + b_{R',12} L^{-2\omega} + 
    b_{R',21} L^{-\omega_2} + \cdots\right)\; .
\label{Rprimeexp}
\end{equation}
In the following section we obtain the estimates
\begin{equation}
    b_{R'_\xi,11}(p=0.65) = 0.60(15), \qquad\qquad b_{U'_4,11}(p=0.65) = 0.40(15).
\label{stime-bRp}
\end{equation}
Bound (\ref{bound-c11}) gives then $|b_{R',11}(p=0.80)| \lesssim 0.02$ 
for both $R'_\xi$ and $U'_4$.
We have thus repeated the analysis at $p=0.8$ considering 
$R'/(1 \pm 0.02 L^{-\omega})$, with $\omega = 0.33$. The results for the exponent $\nu$ 
vary by $\pm 0.0004$, which is negligible with respect to the final error quoted in 
Eq.~(\ref{nuestimate}).

The result (\ref{nuestimate}) is in agreement with and  improves earlier MC estimates:
$\nu=0.683(3)$ (Ref.~\cite{CMPV-03}) and $\nu=0.6837(53)$ 
(Ref.~\cite{BFMMPR-98}). It is also in agreement with the FT result
\cite{PV-00} $\nu=0.678(10)$.

To check universality, it is interesting to compute $\nu$ directly, 
using the data at 
$p = 0.65$. We fitted $\ln R'$ to $a + (1/\nu) \ln L + c L^{-\omega} + 
d  L^{-\epsilon}$ taking $\omega = 0.33(3)$ and $\epsilon\in [0.6,0.9]$ 
(as before the last term takes into account corrections of order 
$L^{-2\omega}$ and $L^{-\omega_2}$). We obtain $\nu = 0.65(2)$, 
$0.67(2)$, $0.68(2)$ for $L_{\rm min} = 8,12,16$. The somewhat large error
is mainly due to the variation of the estimate with $\epsilon$. Thus, if scaling 
corrections are taken into account, universality is satisfied.

\subsection{Improved estimators for the critical exponent $\nu$}
\label{nusecimp}

As we did in Sec.~\ref{uo} for the phenomenological couplings, 
we wish now to define improved estimators of the critical exponent $\nu$, 
i.e., quantities $R'_{\rm im}$ such that $b_{R'_{\rm im},11} = 0$,
see Eq.~(\ref{Rprimeexp}).
We will again combine data at $p=0.8$ with data at $p=0.65$. 

\begin{figure}[tb]
\vspace{1cm}
\centerline{\psfig{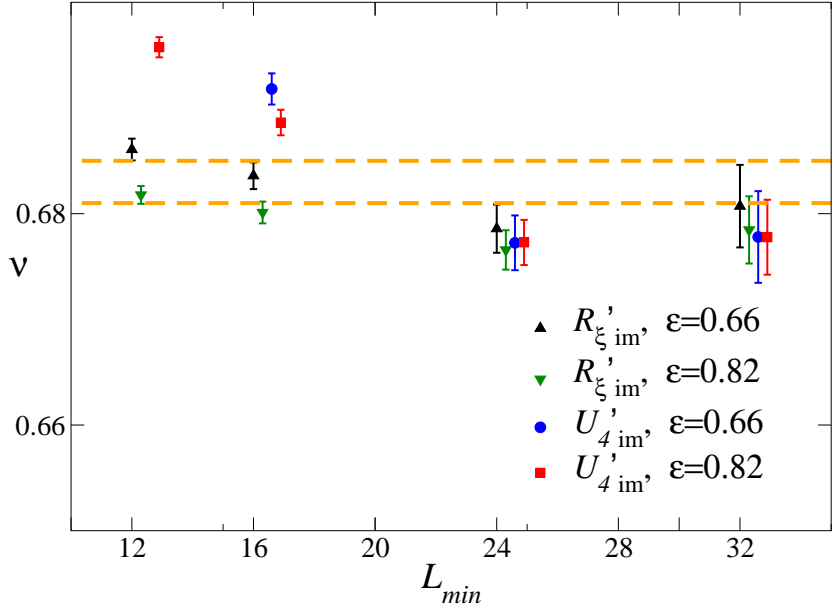}}
\vspace{2mm}
\caption{
Results of fits to $aL^{1/\nu}(1+ c L^{-\epsilon})$ of 
$R'_{\xi,{\rm im}}$ and $U'_{4,{\rm im}}$  for the RSIM at $p=0.65$.
Some results are slightly
shifted along the $x$-axis to make them visible. The dashed lines correspond to 
the estimate $\nu=0.683(2)$ obtained from the analysis of the
data at $p=0.8$. 
}
\label{nup65est}
\end{figure}

Let us consider a phenomenological coupling.  In the following we choose
$\bar{U}_d$ defined in Eq.~(\ref{udiff}) because it has the least relative
statistical errors among the combinations of $\bar{U}_4$ and $\bar{U}_{22}$.
For generic values of $p$ it has the asymptotic behavior
\begin{equation}
\bar{U}_d(p;L) = \bar{U}_d^* \left( 1 + \bar{c}_{d,11}(p) L^{-\omega} + \cdots\right).
\label{cbarp}
\end{equation}
Analyzing the data as described in Secs.~\ref{u22u4}
and \ref{estomega}, we obtain the estimate
$\bar{c}_{d,11}(p=0.65) = - 0.16(2)$. Then, given a generic coupling $R'$
with  asymptotic behavior (\ref{Rprimeexp}),  we have 
\begin{equation}
{R'(p=0.65;L)\over R'(p=0.8;L)}
= A \left( 1 + \Delta b_{R',11} L^{-\omega} + \cdots \right),
\label{rrratio}
\end{equation}
where $\Delta b_{R',11} = b_{R',11}(p=0.65) - b_{R',11}(p=0.8)$. 
A fit of the MC data to Eq.~(\ref{rrratio}) gives
$\Delta b_{R'_\xi,11}=0.60(15)$ and
$\Delta b_{U_4',11}=0.40(15)$.
Bound 
(\ref{bound-c11}) implies $|b_{R',11}(p=0.8)| \lesssim |b_{R',11}(p=0.65)|/30$. 
Therefore, given the error bars on the previous results, we can identify 
$\Delta b_{R',11}$ with $b_{R',11}(p=0.65)$, obtaining the estimates already reported in 
Eq.~(\ref{stime-bRp}). Finally we consider 
\begin{equation}
R'_{\rm im} = 
   R' \bar{U}_d^a \sim L^{1/\nu} [1 + (b_{R',11} + a \bar{c}_{d,11}) L^{-\omega} +
   \cdots ]\; .
\end{equation}
If we fix $a = -b_{R',11}/\bar{c}_{d,11}$, the quantity $R'_{\rm im}$ is improved. 
Using the results obtained above, we find that 
\begin{eqnarray}
&R'_{\xi,{\rm im}} \equiv R'_\xi \bar{U}_d^{a_\xi}\quad {\rm with} \quad a_\xi = 4(1), 
\label{rxipim}\\
&U'_{4,{\rm im}} \equiv U'_4 \bar{U}_d^{a_u} \quad {\rm with} \quad a_u = 2.5(1.0) 
\label{u4pim}
\end{eqnarray}
are approximately improved quantities.  As a check of the results, we perform
the fits (\ref{rrratio}) also for the improved observables. In both cases
$\Delta b_{R',11}$ is consistent with zero.

At $p=0.8$ these improved quantities give results which are substantially
equivalent to those of the original quantities, as shown in Fig.~\ref{nuest},
where we report the estimates corresponding to the central values of the
exponents $a_\xi$ and $a_u$.  This is not unexpected, since the RSIM at
$p=0.8$ has suppressed leading scaling correction for any quantity.

The use of the improved estimators is particularly convenient at $p=0.65$.
Indeed, as discussed in the previous Section, the direct analysis of $R'_\xi$
and $U'_4$ gives estimates of $\nu$ with a large error. In the improved
quantities $R'_{\xi,{\rm im}}$ and $U'_{4,{\rm im}}$ the leading scaling
correction is absent and thus one should be able to determine more precise
estimates of $\nu$ and perform a more severe consistency check of
universality. Since $b_{R',11} = 0$, we fit the data to
\begin{equation}
R'_{\rm im} = a L^{1/\nu} \left( 1 + c L^{-\epsilon} \right)
\label{Rbehfit}
\end{equation}
with $\epsilon \in [0.6,0.9]$.  The results of these fits are shown in
Fig.~\ref{nup65est} (we report results corresponding to the two choices
$\epsilon = 0.66 \approx 2\omega$ and $\epsilon = 0.82 \approx \omega_2$).
Explicitly, we find
$\nu = 0.687(7)$, 0.684(7), 0.679(6), 0.680(8) from the analysis of
$R'_{\xi,{\rm im}}$ with $L_{\rm min} = 12$, 16, 24, 32. The error includes the
statistical error, the error on $a_\xi$, and the variation of the estimate
with $\epsilon$. The results for $U'_{4,\rm im}$ are perfectly consistent. The
estimates are three times more precise than those obtained by using the 
unimproved
$R_\xi'$ and $U_4'$ and are in good agreement with the more precise estimate
$\nu=0.683(2)$ obtained from the data at $p=0.8$.

\subsection{Estimate of the critical exponent $\eta$}
\label{etasec}

\begin{figure}[tb]
\vspace{1cm}
\centerline{\psfig{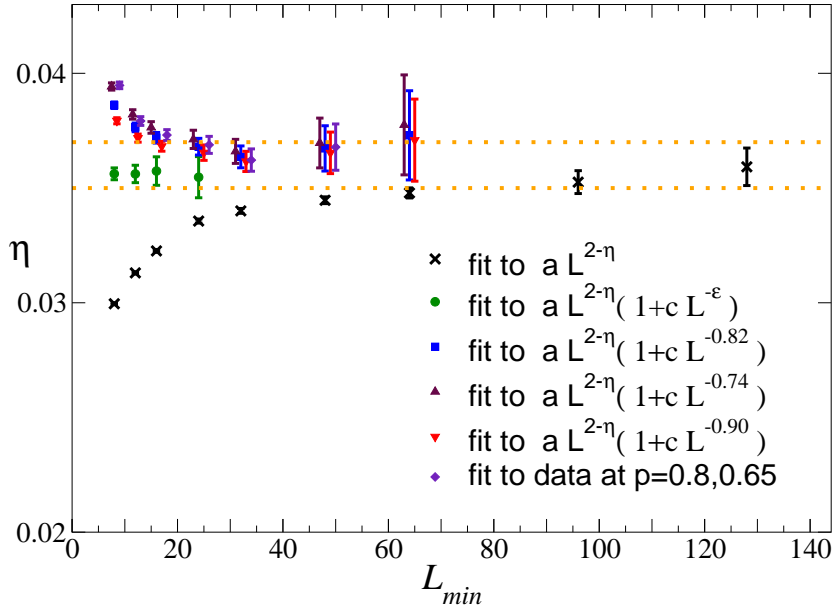}}
\vspace{2mm}
\caption{
Estimates of the critical exponent $\eta$, obtained by fitting the 
magnetic susceptibility $\chi$. All fits refer to the RSIM at $p=0.8$,
except the last one where we simultaneously fit data at $p=0.8$ and at
$p=0.65$ (see text).
Some results are slightly shifted along the $x$-axis to make
them visible. The dotted lines correspond to the final estimate $\eta=0.036(1)$.
}
\label{etaest}
\end{figure}

In order to estimate the critical exponent $\eta$ we analyze the 
magnetic susceptibility $\chi$.  
We have fitted the data at $p=0.8$ to 
$aL^{2-\eta}$, to $a L^{2-\eta} (1 + c L^{-\epsilon})$, taking 
$\epsilon$ as a free parameter, and to $a L^{2-\eta} (1 + c L^{-\omega_2})$ with
$\omega_2=0.82,0.74,0.90$ [we use, as before, the FT estimate 
$\omega_2 = 0.82(8)$]. Moreover, we simultaneously fit the data at
$p=0.8$ and $p=0.65$ to $a_1 L^{2-\eta} (1 + c_1 L^{-0.82})$
if $p=0.8$ and $a_2 L^{2-\eta} (1 + c_2 L^{-0.33})$ if $p=0.65$.
The results are shown in Fig.~\ref{etaest}.
They are fully consistent and provide the final estimate
\begin{equation}
\eta=0.036(1).
\label{etaestimate}
\end{equation}
Again we should discuss the error due to possible residual leading scaling 
corrections at $p=0.8$.
A fit of the data at $p=0.65$ to $a L^{2-\eta} (1 + c
L^{-\omega_2})$ gives $\eta\approx 0.035$. Even if we neglect the correction
term proportional to $L^{-\omega}$, the result of the fit is compatible with 
the estimate (\ref{etaestimate}). This means that the amplitude of 
$L^{-\omega}$ is quite small for $p=0.65$ and gives a
correction of the order of the statistical error in Eq.~(\ref{etaestimate}). 
The amplitude of $L^{-\omega}$ at $p=0.8$  is at least 30 times smaller, 
see Eq.~(\ref{bound-c11}). Therefore, residual scaling corrections are 
negligible. 

The result (\ref{etaestimate}) improves earlier estimates: $\eta=0.035(2)$ and
$\eta=0.0374(45)$ from MC simulations of Refs.~\cite{CMPV-03} and
\cite{BFMMPR-98} respectively. It is also close to, though not 
fully compatible with, the FT result $\eta=0.030(3)$ \cite{PV-00}.

\section{Finite-size scaling analysis of the random site-diluted Ising model
at fixed $\beta$} \label{FSSfixedbeta}

In this section we perform a different analysis using the results at 
$\beta=\beta_{\rm run}$ (the value of $\beta$ at which we performed the 
MC simulation) for the RSIM at $p=0.8$. Thus, the data we use here are 
different from those considered in the previous Section.
We will combine them with those\footnote{
More precisely, the results of Ref.~\cite{CMPV-03} consists in 
5 data with $L=128$ ($N_s = 14000$), 
7 data  with $L=64$ and $N_s = 20000$ and 3 data with $L=64$ and 
$N_s = 40000$, 8 data with $L = 32$ and $N_s = 35000$, and 
4 data with $L = 16$ and $N_s = 80000$. Note that in Ref.~\cite{CMPV-03} 
the derivatives $R'$ were not determined, so that the analysis of 
$\nu$ relies mainly on the present data.} obtained close to the 
critical point ($0.28572\le \beta \le 0.28578$) in Ref.~\cite{CMPV-03}. 
The statistics of the old results is comparable with that obtained here for 
$p=0.8$; note however
that we have now also results for the larger lattice size $L=192$.

\subsection{Phenomenological couplings and critical temperature} 
\label{anaR-fixedb}

\begin{table}
\footnotesize
\begin{center}
\begin{tabular}{rlllll}
\hline
$L_{\rm min}$ & \multicolumn{1}{c}{$\beta_c$} & 
\multicolumn{1}{c}{$U_4^*$} & 
\multicolumn{1}{c}{$U_{22}^*$} & 
\multicolumn{1}{c}{$R_\xi^*$} & 
\multicolumn{1}{c}{$\epsilon$} \\
\hline
 8&   0.2857432(3) & 1.6476(10) & 0.14779(26) & 0.59394(23) & 0.97(7)  \\
12&   0.2857432(3) & 1.6480(15) & 0.14791(32) & 0.59393(30) & 0.94(13) \\
16&   0.2857430(3) & 1.6480(18) & 0.14787(35) & 0.59363(32) & 0.97(18) \\
24&   0.2857434(4) & 1.6471(27) & 0.14801(51) & 0.59427(66) & 0.99(45)  \\
\hline
8  & 0.2857430(3)[1] & 1.6499(5)[16]&0.14761(29)[13]& 0.59410(26)[11] & fixed \\
12 & 0.2857430(3)[1] & 1.6494(6)[13]&0.14781(33)[9] & 0.59403(31)[8]  & fixed \\
16 & 0.2857429(3)[1] & 1.6497(7)[12]&0.14775(37)[9] & 0.59370(36)[5]  & fixed \\
24 & 0.2857433(4)[1] & 1.6482(12)[7]&0.14793(55)[5] & 0.59444(64)[10] & fixed \\
32 & 0.2857436(5)[1] & 1.6472(14)[6]&0.14768(61)[7] & 0.59494(74)[14] & fixed \\
\hline
\end{tabular}
\end{center}
\caption{Results of the combined fits for the phenomenological couplings 
for several values of $L_{\rm min}$, the smallest lattice size used in the 
analyses. In the first set of fits, $\epsilon$ is free, while in the 
second set we fixed $\epsilon=\omega_2 = 0.82(8)$. The number in parentheses
is the statistical error, while the number in brackets gives the 
variation of the estimate as $\omega_2$ is varied within one error bar.}
\label{table-Rfixedb}
\end{table}

As discussed in Sec.~\ref{FSS-general}, close to the critical point
a phenomenological coupling behaves as [see Eq.~(\ref{expandR})]
\begin{equation}
R(\beta,L) = R^* + a_{1} (\beta - \beta_c) L^{1/\nu} + 
          a_{2} L^{-\epsilon}.
\label{fitR-fixedb}
\end{equation}
In order to determine $R^*$ and $\beta_c$ we have performed two different types of fit,
always fixing $\nu = 0.683(5)$ [results are essentially unchanged if we vary
$\nu$ in $[0.678,0.688]$, which is quite conservative given the estimate 
(\ref{nuestimate})]. In the first case, we simultaneously fit $R_\xi$, $U_4$, and 
$U_{22}$, keeping $\epsilon$ as a free parameter. For the exponent $\epsilon$
we obtain $\epsilon = 0.95(20)$: as expected there is no indication of 
a correction-to-scaling term with exponent $\omega \approx 0.33$. 
In the second fit, we use the fact that the model is improved, 
set $\epsilon = \omega_2$,
and use the FT estimate $\omega_2 = 0.82(8)$. The results of the fits 
for several values of $L_{\rm min}$ are reported in Table
\ref{table-Rfixedb}. They are quite stable with respect to $L_{\rm min}$ and indeed
results with $L_{\rm min} = 8$ are compatible with all those that correspond to 
larger values. If we take conservatively the final estimates from the 
results with $L_{\rm min} = 24$ and $\omega_2$ fixed, we have
\begin{eqnarray}
\beta_c &=& 0.2857433(5), \nonumber \\
R^*_\xi &=& 0.5944(7), \nonumber \\
U^*_4   &=& 1.648(2), \nonumber \\
U^*_{22}&=& 0.1479(6) .
\end{eqnarray}
The estimate of $\beta_c$ is compatible with that reported in Sec.~\ref{betacRSIM},
$\beta_c = 0.2857429(4)$. The estimate of $R^*_\xi$ is essentially identical to that 
reported in Ref.~\cite{CMPV-03}, $R^*_\xi = 0.5943(9)$, so that our analysis 
at fixed $R_\xi = 0.5943$ corresponds indeed to fixing $R_\xi = R^*_\xi$. 
This is also confirmed by the results for $U^*_4$ and $U^*_{22}$ that are compatible
with the estimates of $\bar{U}^*_4$ and $\bar{U}^*_{22}$ obtained in 
Sec.~\ref{u22u4} (if we had performed analyses at fixed $R_\xi \not=R^*_\xi$
such an equality would not hold, see Sec.~\ref{sec3.2}). The estimates of 
${U}^*_4$ and ${U}^*_{22}$ agree with previous MC estimates:
$U^*_4 = 1.650(9)$ and $U^*_{22}= 0.1480(10)$ (Ref.~\cite{CMPV-03});
$U^*_4 = 1.653(20)$ and $U^*_{22}= 0.145(7)$  (Ref.~\cite{BFMMPR-98}).

\subsection{Estimates of $\nu$} \label{nu-fixedb}

\begin{figure}[tb]
\vspace{1cm}
\centerline{\psfig{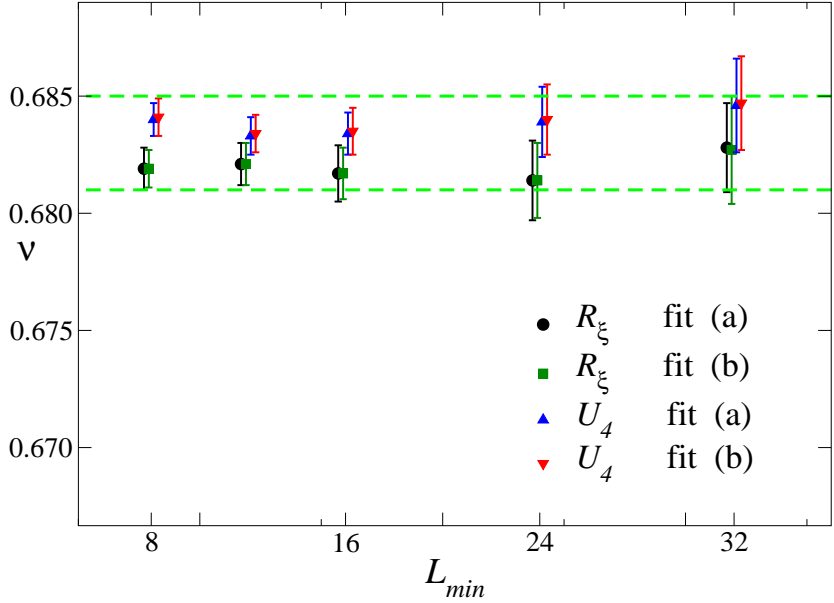}}
\vspace{2mm}
\caption{
Estimates of the critical exponent $\nu$, obtained by simultaneous fits of 
$R_\xi$ and $R'_\xi$, and of $U_4$ and $U'_4$ (see text).
Some results are slightly shifted along the $x$-axis to make
them visible. The dashed lines correspond to the estimate $\nu=0.683(2)$ 
obtained in Sec.~\protect\ref{nusec}.
}
\label{fig:nu-fixedb}
\end{figure}

We now compute the critical exponent $\nu$. It may be obtained by 
fitting $R'(\beta=\beta_c,L)$ to $a L^{1/\nu}$. This requires fixing 
$\beta_c$ and this induces a somewhat large error. We have found more convenient
to follow a different route, analyzing simultaneously $R'$ (this gives $\nu$)
and $R$ (this essentially fixes $\beta_c$). We performed two types of fits. 
First [fit (a)], we fit $R$ and $R'$ to 
\begin{eqnarray}
R(\beta,L) &=& R^* + a_{1} (\beta - \beta_c) L^{1/\nu} + a_{2} L^{-\epsilon} 
\nonumber \\
&& \qquad 
 + a_{3} (\beta - \beta_c)^2 L^{2/\nu} + 
   a_{4} (\beta - \beta_c) L^{1/\nu-\epsilon}, \nonumber \\
R'(\beta,L) &=& a_{1} L^{1/\nu} + 
 2 a_{3} (\beta - \beta_c)  L^{2/\nu} + 
 a_{4} L^{1/\nu-\epsilon}.
\end{eqnarray}
Here $\beta_c$ and $\nu$ are kept as free parameters, while $\epsilon$ is 
fixed: $\epsilon=\omega_2 = 0.82(8)$. In the second fit [fit (b)], 
we fit $R$ and $R'$ to
\begin{eqnarray}
&& R(\beta,L) = R^* + a_{1} (\beta - \beta_c) L^{1/\nu} + a_{2} L^{-\epsilon} ,
\nonumber \\
&& \ln [R'(\beta,L)] = a_{3} + {1\over \nu} \ln L + 
 a_{4} (\beta - \beta_c)  L^{1/\nu} + 
 a_{5} L^{-\epsilon},
\end{eqnarray}
where again $\epsilon=\omega_2 = 0.82(8)$.  The two fits give similar results,
see Fig.~\ref{fig:nu-fixedb}. For instance, for $L_{\rm min} = 24$ and $R =
R_\xi$ we have $\nu = 0.6814(17)$ and 0.6814(16) from fit (a) and (b),
respectively (the reported errors are the sum of the statistical error and of
the variation of the estimate as $\omega_2$ is varied within one error bar).
If $R = U_4$ we obtain analogously $\nu = 0.6839(15)$ and 0.6840(16).
Collecting results, this type of analysis gives the final estimate
\begin{equation}
\nu = 0.6825(25),
\end{equation}
which is fully compatible with Eq.~(\ref{nuestimate}).

\subsection{Estimates of $\eta$} 

\begin{figure}[tb]
\vspace{1cm}
\centerline{\psfig{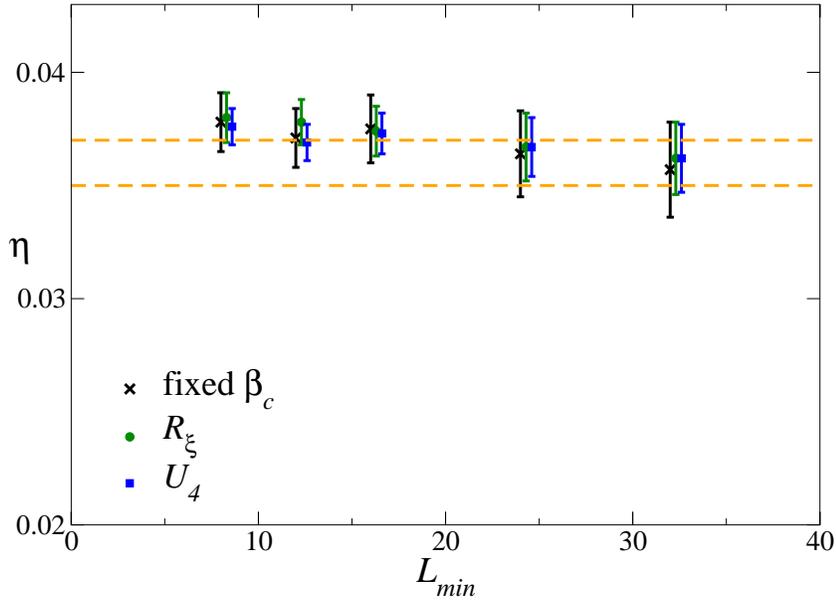}}
\vspace{2mm}
\caption{
Estimates of the critical exponent $\eta$, obtained in different fits 
(see text).
Some results are slightly shifted along the $x$-axis to make
them visible. The dashed lines correspond to the estimate 
$\eta=0.036(1)$ obtained in Sec.~\protect\ref{etasec}.
}
\label{fig:eta-fixedb}
\end{figure}

As in Ref.~\cite{CMPV-03} we determine $\eta$ from the critical behavior of 
$Z \equiv  \chi/\xi^2$, which is more precise than $\chi$: 
The relative error on $\chi$
is 3.4 times larger than the relative error on $Z$. 
We perform two types of fits.
First, we analyze $Z$ as 
\begin{equation}
\ln [Z(\beta,L)]
 = a - \eta \ln L + b L^{1/\nu} (\beta - \beta_c) + c L^{-\omega_2},
\end{equation}
fixing $\nu$, $\omega_2$, and $\beta_c$. The estimates are little sensitive to
$\nu$ and $\omega_2$ that are fixed to $\nu = 0.683(5)$ and $\omega_2 = 0.82(8)$. 
The dependence on $\beta_c$ is instead significant, of the order of the 
statistical error. We use
$\beta_c = 0.2857431(6)$ that combines the estimates determined in 
Sec.~\ref{betacRSIM} and \ref{anaR-fixedb}. Results are reported in 
Fig.~\ref{fig:eta-fixedb}.
For $L_{\rm min} = 24$, $\eta = 0.0364(9+2+8)$ where we quote the 
statistical error, the variation of the estimate with $\omega$ and $\nu$,
and the change as $\beta_c$ varies within one error bar. 

As done in Sec.~\ref{nu-fixedb}, we can avoid using $\beta_c$ by analyzing 
$\ln Z$ together with a renormalized coupling $R$. We fit $R$ to 
Eq.~(\ref{fitR-fixedb}), again fixing $\nu$ and $\epsilon = \omega_2$. The results 
are reported in Fig.~\ref{fig:eta-fixedb}.
For $L_{\rm min} = 24$ we obtain 
$\eta = 0.0367(15)$ and $\eta = 0.0367(13)$ using $R_\xi$ and 
$U_4$, respectively. Collecting results this analysis provides the 
final estimate 
\begin{equation}
\eta = 0.0365(15),
\end{equation}
which agrees with
the estimate $\eta = 0.036(1) $ obtained in the analyses at fixed $R_\xi$.

\section{Finite-size scaling analysis of the randomly bond-diluted Ising model} 
\label{FSSRBIM}

\begin{figure}[tb]
\vspace{1cm}
\centerline{\psfig{width=8.2truecm,angle=0,file=u22-bp55.eps}}
\vspace{0.5cm}
\centerline{\psfig{width=8.2truecm,angle=0,file=ud-bp55.eps}}
\vspace{0.0cm}\hspace{0.1cm} 
\centerline{\psfig{width=6.5truecm,angle=-90,file=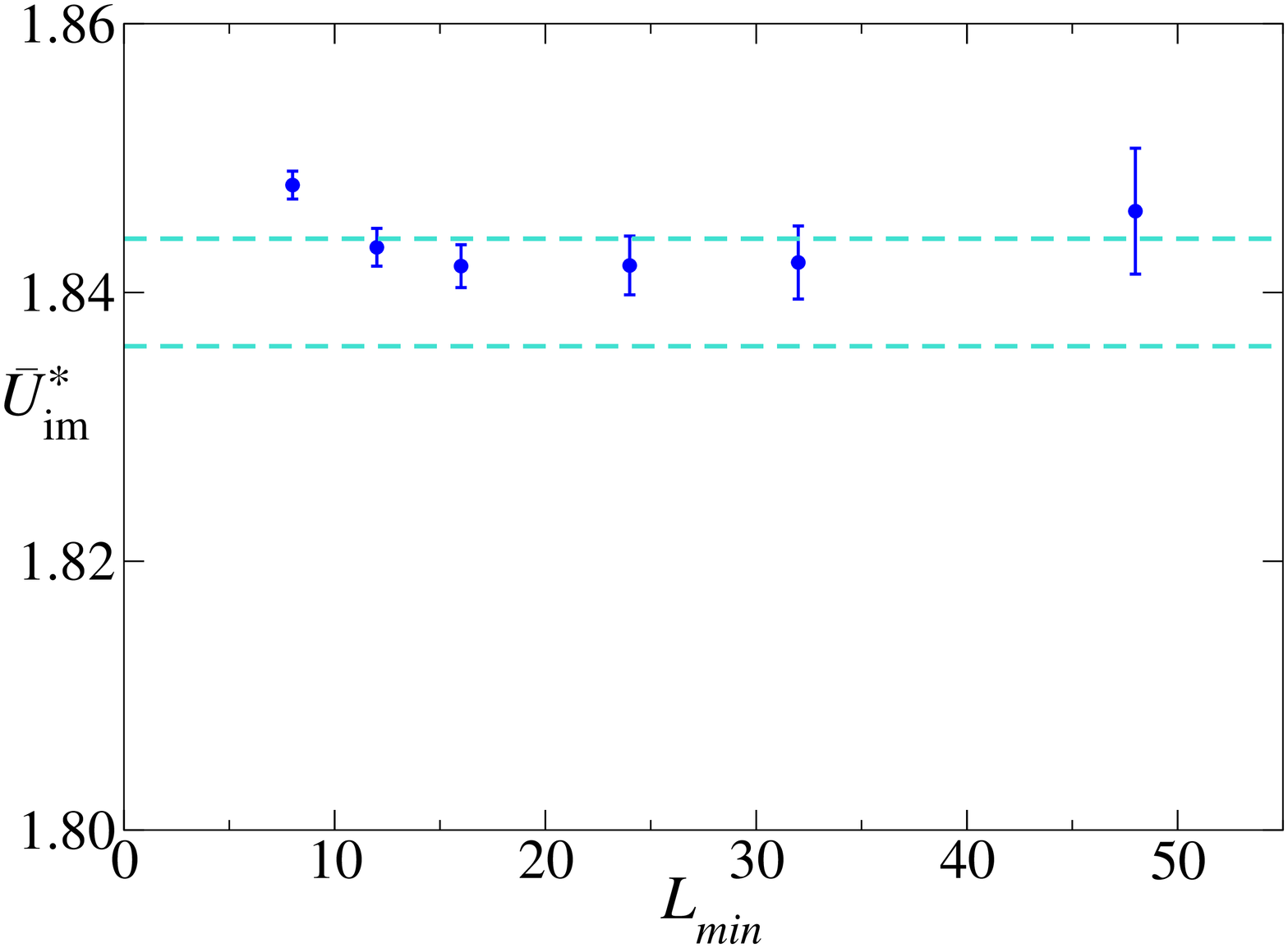}}
\vspace{2mm}
\caption{
Estimates of $\bar{U}_{22}^*$,
$\bar{U}_{\rm im}^*$, and $\bar{U}_{d}^*$ obtained by 
fitting $\bar{U}_{22}$,
$\bar{U}_{\rm im}$, and $\bar{U}_{d}$ for the RBIM at $p=0.55$,
to $\bar{U}^* + c L^{-\omega_2}$ with $\omega_2=0.82$.
The dashed lines correspond to the estimates of $\bar{U}^*$ 
obtained from the analyses of the data of the RSIM at $p=0.8$.
}
\label{u-bp55}
\end{figure}

In this section we analyze the MC data of the RBIM at
$p=0.55$ and $p=0.7$ at fixed $R_\xi=0.5943$. They are reported in 
Tables~\ref{tablerun0.55} and \ref{tablerun0.7}, respectively.  We show that
their critical behavior is fully consistent with 
that obtained for the RSIM in Secs.~\ref{FSSRSIM} and \ref{FSSfixedbeta}. 

\subsection{Phenomenological couplings}

We first discuss the FSS behavior of the quartic cumulants. 
In Fig.~\ref{u22omega} we have already shown the MC results
$\bar{U}_{22}(L)$ for the RBIM at $p=0.55$ and $p=0.7$ 
versus $L^{-\omega}$ with $\omega=0.33$. 
Their large-$L$ behavior is perfectly
consistent with that observed in the RSIM, all data converging to 
$\bar{U}_{22}^*=0.148(1)$ as $L$ increases. A more quantitative check 
can be performed by fitting the MC results for $\bar{U}_{\rm im}$.
Indeed, as discussed in Sec.~\ref{uo}, the leading scaling corrections
in $\bar{U}_{\rm im}$ are at least a factor of 10 and a factor of 20 smaller that those 
in $\bar{U}_{22}$ and $\bar{U}_{d}$, respectively.
Thus, for any generic $p$, 
we should be able to obtain estimates of $\bar{U}_{\rm im}^*$ that are more 
precise than those of $\bar{U}_{22}^*$ and $\bar{U}_{d}^*$. In particular, we 
expect errors comparable to that of the RSIM result 
$\bar{U}_{\rm im}^* = 1.840(4)$. We fit $\bar{U}_{\rm im}$ to 
$\bar{U}_{\rm im}^* + c L^{-\epsilon}$ with $\epsilon = 0.66$ and 0.82
(we remind the reader that the leading scaling corrections to $\bar{U}_{\rm im}$
are proportional to $L^{-2\omega}$ and $L^{-\omega_2})$. Results are 
reported in Fig.~\ref{uimfits}. They are fully compatible with those obtained in the 
RSIM at $p = 0.8$, confirming that the RBIM belongs to the same 
universality class of the RSIM.

A direct analysis of $\bar{U}_{22}$ and $\bar{U}_d$ for the RBIM at $p=0.7$
gives results with large errors (see the corresponding analysis for the 
RSIM at $p=0.65$ reported in Sec.~\ref{estomega}). Universality is verified,
though with limited precision. More precise results are obtained for the 
RBIM at $p=0.55$, since for this value of $p$ 
the RBIM  turns out to be approximately improved. To determine $p^*$ for the 
RBIM we follow the same strategy employed in Sec.~\ref{pstar}. We determine 
the correction-to-scaling amplitude $c_{22,11}$ 
obtaining $c_{22,11}(p=0.55)=-0.01(2)$ and $c_{22,11}(p=0.7) = -0.17(2)$. 
Then, we assume that for $p\approx 0.55$, 
$c_{11}(p)\approx a (p-p^*)$, so that 
\begin{equation}
p^* = 0.55 - {1\over a} c_{22,11}(p=0.55).
\end{equation}
The constant $a$---only a rough estimate is needed---is again determined as 
\begin{equation}
a = \left. {dc_{22,11}\over dp}\right|_{p=p^*} \approx
  {c_{22,11}(p=0.7) - c_{22,11}(p=0.55) \over 0.7 - 0.55} = -1.07(15).
\label{detaRBIM}
\end{equation}
This gives 
\begin{equation} 
p^*=0.54(2)
\end{equation}
for the RBIM. Again, in setting the error we have not considered the 
error on the linear interpolation (\ref{detaRBIM}).

Since the RBIM at
$p=0.55$ is approximately improved, we can fit $\bar{U}_{22}$,
$\bar{U}_{\rm im}$, and $\bar{U}_{d}$ to $\bar{U}^* + c
L^{-\omega_2}$.  In Fig.~\ref{u-bp55} we show the results for
$\omega_2=0.82$.  They are very stable and in perfect agreement with
the results obtained from the FSS analysis of the RSIM at $p=0.8$. Indeed, we
obtain
\begin{equation}
\bar{U}_{\rm im}^*=1.842(4),\quad
\bar{U}_{22}^*=0.148(1),\quad
\bar{U}_{d}^*=1.501(2),
\label{uestbp55}
\end{equation}
where the error takes into account the uncertainty on $\omega_2$.
They must be compared with the estimates obtained from the FSS
analysis of the RSIM: $\bar{U}_{\rm im}^*=1.840(4)$,
$\bar{U}_{22}^*=0.148(1)$, and $\bar{U}_{d}^*=1.500(1)$. These results
provide strong evidence for universality between the RSIM and the RBIM.

\subsection{Critical temperatures}

Let us now estimate $\beta_c$ from the estimates of $\beta_f$.  For $p =
0.55$, since the model is approximately improved, we can fit the data of
$\beta_f$ to $\beta_c+cL^{-1/\nu-\omega_2}$.  We obtain
$\beta_c=0.4322895(15)$, which is compatible with the MC results of
Ref.~\cite{BCBJ-04}, $\beta_c=0.43225(10)$.  The analysis of the 19th-order
high-temperature expansion of $\chi$ reported in Ref.~\cite{JBCBH-05} gave the
estimate $\beta_c=0.43253(12)$.

For $p=0.7$ we must take into account the leading scaling corrections, 
i.e. fit $\beta_f$ to $\beta_c+cL^{-\epsilon}$,
where $\epsilon \in [\omega + 1/\nu,\omega_2 + 1/\nu]$. 
We obtain $\beta_c = 0.326707(2)$,
which agrees with the MC estimate 
$\beta_c= 0.32670(5)$ of Ref.~\cite{BCBJ-04}.

\subsection{Critical exponents}

\begin{figure}[!tb]
\vspace{1cm}
\centerline{\psfig{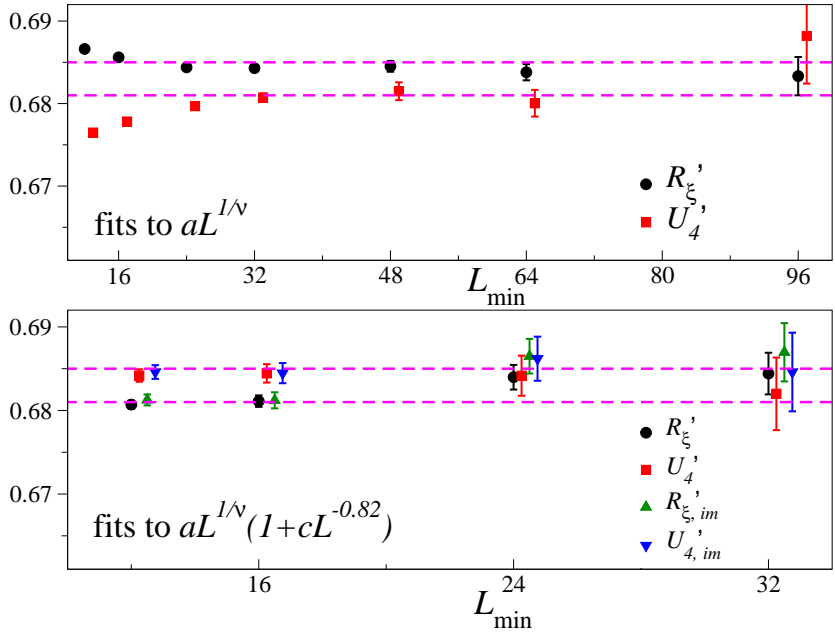}}
\vspace{2mm}
\caption{
Estimates of the critical exponent $\nu$, obtained by fitting the data
of the RBIM at $p=0.55$ 
(some results are slightly shifted along the
$x$-axis to make them visible).  The dashed lines correspond to the 
estimate $\nu=0.683(2)$ obtained in the RSIM at $p=0.8$.  }
\label{nubp55}
\end{figure}

\begin{figure}[tbh]
\vspace{1cm}
\centerline{\psfig{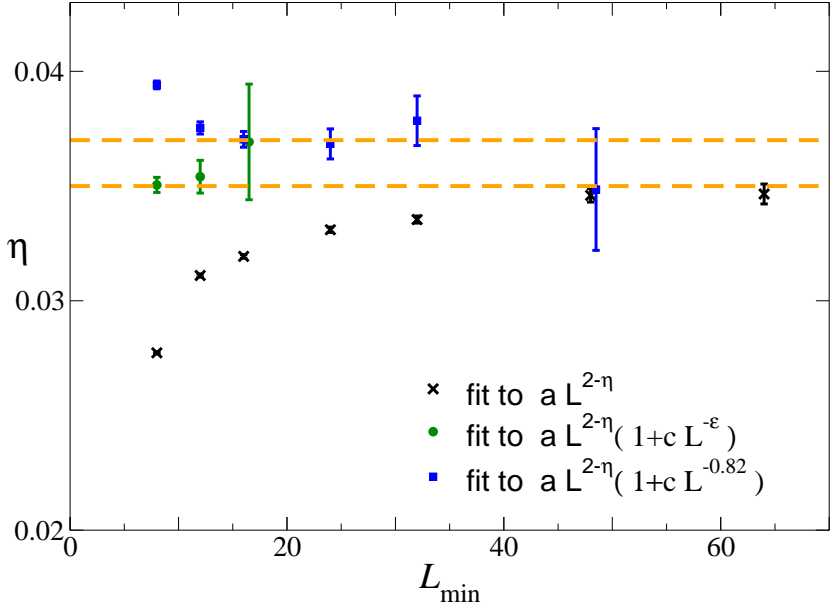}}
\vspace{2mm}
\caption{
Estimates of the critical exponent $\eta$, obtained by fitting the 
data of the RBIM at $p=0.55$.
The dashes line correspond to the estimate $\eta=0.036(1)$
obtained in the RSIM at $p=0.8$.
}
\label{etabp55}
\end{figure}

\begin{figure}[tb]
\vspace{1cm}
\centerline{\psfig{width=11truecm,angle=0,file=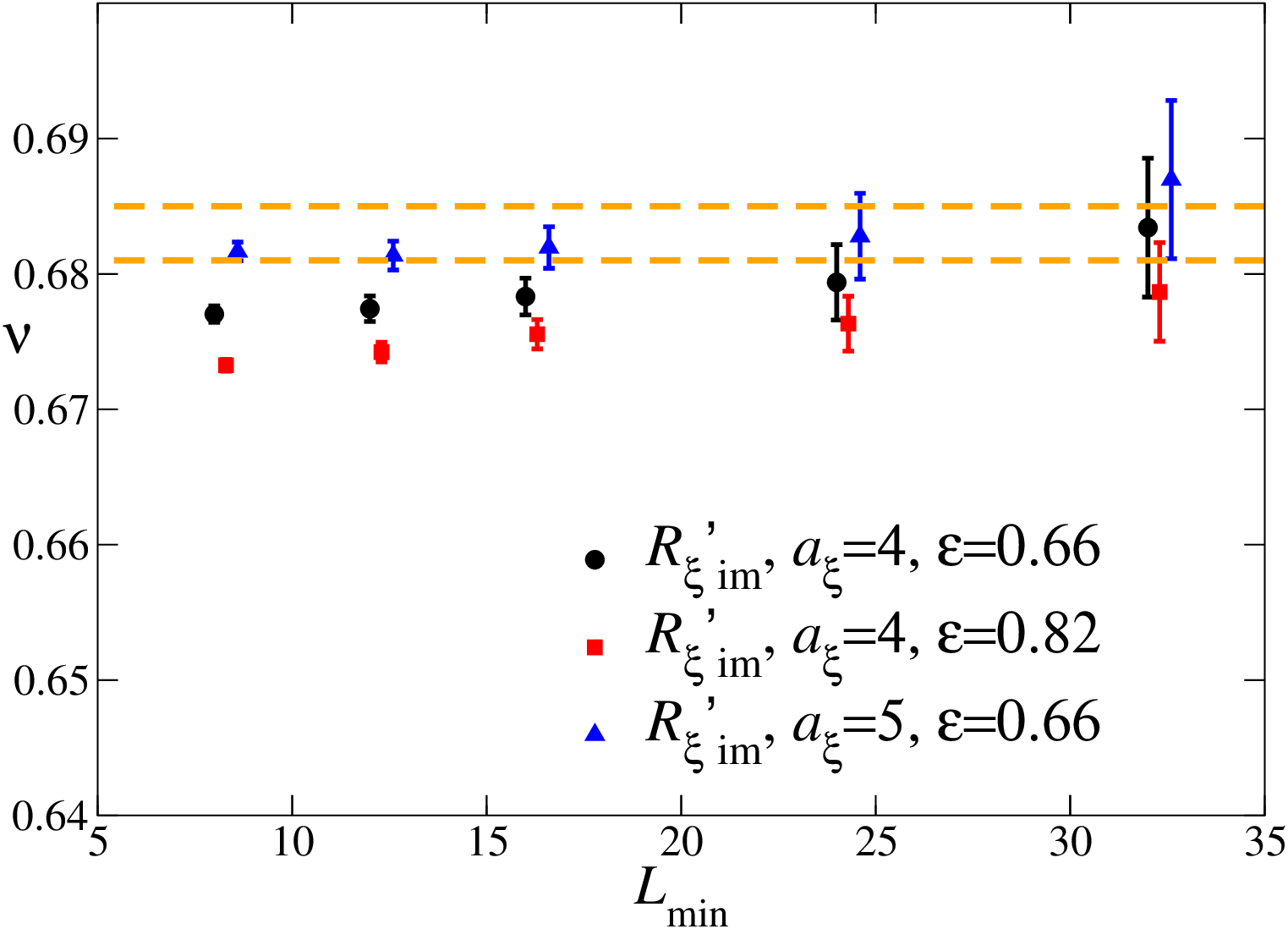}}
\vspace{2mm}
\caption{
Results for the critical exponent $\nu$, obtained by fitting the 
data of the RBIM at $p=0.7$
(some results are slightly shifted along the
$x$-axis to make them visible).  The dashed lines correspond to the 
estimate $\nu=0.683(2)$ obtained in the RSIM at $p=0.8$.  
}
\label{nubp7}
\end{figure}

Let us now consider the critical exponents.  Since the RBIM at $p=0.55$ is
approximately improved, we perform the same analysis as done for the RSIM 
at $p=0.8$, see Sec.~\ref{nusec}.
In Fig.~\ref{nubp55} we show the results of several fits of the data of the
derivative of the phenomenological couplings at $p=0.55$.  The estimates obtained 
by using $R'_\xi$ and $R'_{\xi, {\rm im}}$ are substantially equivalent, as
expected because the RBIM at $p=0.55$ is approximately improved.  The estimates 
of $\nu$ shown in Fig.~\ref{nubp55} are fully consistent with
the estimate $\nu=0.683(2)$ obtained from the FSS analysis of the RSIM at
$p=0.8$.  
Similar conclusions hold for the critical exponent $\eta$. In
Fig.~\ref{etabp55} we show the results of several fits analogous to those 
discussed in Sec.~\ref{etasec}. Again universality is well satisfied:
the estimates of $\eta$ are compatible with the RSIM result 
$\eta = 0.036(1)$. 

In the case of the RBIM at $p=0.7$, analyses of unimproved quantities 
give estimates of $\nu$  with large errors. We therefore only consider 
the improved quantities.
In Fig.~\ref{nubp7} we show the results of fits of $R'_{\xi, {\rm im}}$ 
for the RBIM at $p=0.7$. They are again substantially
consistent with the estimate $\nu=0.683(2)$ obtained from the FSS analysis of
the RSIM at $p=0.8$. The MC estimates of $U'_{4, {\rm im}}$ are less 
precise, but  again consistent with universality.

\section*{Acknowledgments}

The  MC simulations have been done at the 
theory cluster of CNAF (Bologna) and at the INFN Computer Center in Pisa.

\appendix

\section{Field-theory estimate of $\omega_2$}
\label{omega2}

In this appendix we estimate $\omega_2$ by a reanalysis
of its FT six-loop expansion in the massive zero-momentum scheme.
In Ref.~\cite{PV-00} we performed a direct analysis of the stability
matrix at the RDIs fixed point (FP), obtaining $\omega_2 = 0.8(2)$. 
Here we shall use the method discussed in Ref.~\cite{CMPV-03}---it consists in an 
expansion around the unstable Ising FP---which
allows us to estimate accurately the 
difference $\omega_2-\omega_{\rm Is}$, where $\omega_{\rm
Is}$ is the leading correction-to-scaling
exponent in the standard Ising model. Since $\omega_{\rm Is}$ 
is known quite precisely, this method allows us to obtain 
a precise result for $\omega_2$.

In the FT approach one starts from Hamiltonian (\ref{Hphi4rim}), determining
perturbative expansions in powers of 
the renormalized couplings $u$ and $v$. We normalize them so that $u\approx u_0$ and 
$v\approx v_0$ at tree level (these are the normalizations used in Ref.~\cite{CMPV-03};
they differ from those of Ref.~\cite{PV-00}). As discussed in 
Ref.~\cite{CPPV-04}, it is more convenient to introduce new variables 
$y\equiv u+v$ and $z\equiv -u$.
The Ising FP is located at $z_I^*=0$ and $y_I^*=g_{\rm
Is}^*$, where \cite{CPRV-02} $g_{\rm Is}^*=23.56(2)$.  
The RDIs FP is located at
$y^*=24.7(2)$ and $z^*= 18.6(3)$ (we use the MC results
of Ref.~\cite{CMPV-03} since the FT estimates $y^*=24.8(6)$ and $z^*= 14(2)$ 
are less precise).

The expansion around the Ising FP can be performed along the
Ising-to-RDIs RG trajectory \cite{CPPV-04}, which is obtained as the
limit $z_0\rightarrow 0^+$ ($u_0\rightarrow 0^-$) 
of the RG trajectories in the $z$, $y$
plane, see Fig.~\ref{figtraj}.  An effective parametrization of the curve is given by the
first few terms of its expansion around $z=0$, which is given by
\begin{equation}
y - y_I = T(z) = c_2 z^2 + c_3 z^3 + \cdots
\label{expansion-T}
\end{equation}
where \cite{CPPV-04} $c_2=0.0033(1)$ and $c_3=1(2)\times 10^{-5}$.
The fact that $y-y_I$ is of order $z^2$ is the main reason why 
the variable $y$ was introduced and is due to the identity 
\cite{CPPV-04}
\begin{eqnarray}
\left. {\partial \beta_v\over \partial u} \right|_{u=0}
+ \left. {\partial \beta_u\over \partial u} \right|_{u=0}
- \left. {\partial \beta_v\over \partial v} \right|_{u=0} =0,
\end{eqnarray}
which corresponds to 
\begin{eqnarray}
\left. {\partial \beta_y\over \partial z} \right|_{z=0} = 0.
\label{betaeq}
\end{eqnarray}
Performing the variable change $y=g+T(z)$  in the double expansion
of a generic quantity $f(y,z)$ in powers of $y$ and $z$, we obtain
\begin{equation}
\bar{f}(g,z)=f(g+T({z}),{z})= \sum_i e_i(g) {z}^i.
\label{eidef}
\end{equation}
The coefficients $e_i$ must be evaluated at $g = g_{\rm Is}^*$.
This is done by resumming their perturbative expansions
as discussed in Ref.~\cite{CPV-00}: in particular,
we exploit Borel summability  and the knowledge of the large-order
behavior at the Ising FP.

In Ref.~\cite{CMPV-03} this approach was applied to the standard
critical exponents.  Here we extend these calculations to the
next-to-leading scaling-correction exponent $\omega_2$. For this
purpose, we consider the stability matrix
\begin{equation}
\Omega = (\partial \beta_{y}/ \partial y, \partial \beta_{y}/ \partial z;
\partial \beta_{z}/ \partial y, \partial \beta_{z}/ \partial z).
\end{equation}
Each entry has an expansion of the form (\ref{eidef}), with coefficients 
$e_i(g)$ that are resummed as discussed before.
Then, the matrix $\Omega$ is diagonalized, obtaining
the expansion of its eigenvalues in powers of $z$ up to
$O(z^3)$. The corresponding coefficients for the smallest
eigenvalue are quite large, so that this method does not provide 
an estimate of $\omega$ which is more precise than that obtained in 
Ref.~\cite{PV-00}, i.e. $\omega=0.25(10)$.
On the other hand, the expansion coefficients for $\omega_2$ are quite small.
To order $z^3$ we obtain 
\begin{eqnarray}
&&\omega_2 - \omega_{\rm Is}= \sum_i c_i z^i, \label{A7} \\
&&c_1 = 0, \quad c_2 = 5(2)\times 10^{-5},  \quad c_3 = - 2(6) \times 10^{-6},
\end{eqnarray}
where $c_1=0$ exactly [this is a consequence of relation (\ref{betaeq})], 
and the errors on the
coefficients $c_2$ and $c_3$ are due to the resummation of the
corresponding series evaluated at $g_{\rm Is}^*$.  
Expansion (\ref{A7}) is evaluated at $z = 18.6(2)$, obtaining
\begin{equation}
\omega_2-\omega_{\rm Is}=0.00(5),
\label{diffomega2}
\end{equation}
where the error takes into account all possible sources of uncertainties:
the error on the coefficients, the truncation of the series in
powers of $z$, and the uncertainty on the coordinates of the RDIs FP.
Then, using the estimate $\omega_{\rm Is}=0.82(3)$, which takes into
account the results of Refs.~\cite{GZ-98,Has-99,DB-03}
obtained by various approaches, we arrive at the estimate
\begin{equation}
\omega_2=0.82(8),
\end{equation}
which improves the result $\omega_2=0.8(2)$ obtained in Refs.~\cite{PV-00,CPPV-04}.

\section{Bias corrections}
\label{appendix_bias}

In this section we discuss the problem of the bias correction needed in 
the calculations of 
disorder averages of combinations of thermal averages. As already emphasized in 
Ref.~\cite{BFMMPR-98-b}, this is a crucial step in high-precision 
MC studies of random systems.

To discuss it in full generality, let us indicate with $S$ the state space 
corresponding to the variables $\sigma$ and with $R$ that corresponding to the 
dilution variables. Then, we consider a probability function $\pi(\sigma;\rho)$
on $S$ depending parametrically on $\rho$ ($\pi = e^{-\beta {\cal H}}/Z$ in the 
specific calculation) and a probability $p(\rho)$ on $R$. Averages over 
$\pi(\sigma;\rho)$ are indicated as $\< \cdot \>$, or with 
$\< \cdot \>_\rho$ when we wish to specify the value $\rho$ used 
in the calculation. 
Moreover, we assume $R$ to have a finite number $K$ of elements 
($K = 2^V$ and $K = 2^{d V}$ in the RSIM and in the RBIM, respectively).
We wish to compute averages of functions $A(\sigma,\rho)$.
We first discuss the calculation of 
\begin{equation}
{\cal O}_n \equiv  \overline{\< A \>^n} = 
  \sum_\rho p(\rho) \left[ \sum_\sigma \pi(\sigma;\rho) A(\sigma,\rho)\right]^n.
\label{defOn}
\end{equation}
A numerical strategy to compute ${\cal O}_n$ could be the following. 
Extract $N_s$ independent disorder configurations $\rho_\alpha$,
$\alpha=1,\ldots,N_s$, with probability $p(\rho)$ and then, for each 
$\rho_\alpha$, extract $N_m$ {\em independent} configurations $\sigma_{a,\alpha}$,
$a = 1,\ldots,N_m$, with probability $\pi(\sigma;\rho_\alpha)$. 
Then, define the sample average
\begin{equation}
[A]_{\rho_\alpha} \equiv  {1\over N_m} \sum_{a=1}^{N_m} 
    A(\sigma_{a,\alpha},\rho_\alpha).
\end{equation}
A possible estimator of ${\cal O}_n$ could be 
\begin{equation}
{\cal O}_n^{\rm est} \equiv  {1\over N_s} \sum_{\alpha=1}^{N_s}
   [A]_{\rho_\alpha}^n.
\label{Onest}
\end{equation}
The question is whether ${\cal O}_n^{\rm est}$ converges to ${\cal O}_n$ 
defined in Eq.~(\ref{defOn}) as $N_s\to\infty$ at {\em fixed} $N_m$. 

To answer this question, let $N(\rho)$ be the number of $\rho_\alpha$ such that 
$\rho_\alpha = \rho$. Eq.~(\ref{Onest}) can thus be rewritten as 
\begin{equation}
{\cal O}_n^{\rm est} = {1\over N_s} \sum_{\rho\in R} 
   \sum_{\alpha=1}^{N(\rho)} [A]_{\rho,\alpha}^n,
\end{equation}
where the second sum extends over the $N(\rho)$ terms that appear in 
Eq.~(\ref{Onest}) such that $\rho_\alpha = \rho$, i.e., which 
correspond to the same 
disorder configuration. As $N_s\to \infty$, 
$N(\rho)$ converges to $N_s p(\rho)$ with probability 1; thus, as $N_s\to \infty$
\begin{equation}
\sum_{\alpha=1}^{N(\rho)} [A]_{\rho,\alpha}^n \to 
N_s p(\rho) \times {1\over N(\rho)} 
        \sum_{\alpha=1}^{N(\rho)} [A]_{\rho,\alpha}^n \to 
   N_s p(\rho) \left\< [A]_{\rho}^n \right\>_\rho.
\end{equation}
Thus, for $N_s\to \infty$ we have 
\begin{equation}
{\cal O}_n^{\rm est} = \sum_\rho p(\rho) \left\< [A]_{\rho}^n \right\>_\rho 
   = \overline{ \left\< [A]^n\right\> }.
\label{Onest2}
\end{equation}
Eq.~(\ref{Onest2}) relies only on the limit $N_s\to \infty$ and is valid for 
any $N_m$. If $n=1$ we obtain 
\begin{equation}
\left\< [A]\right\>_\rho = {1\over N_m} 
   \left \< \sum_{a=1}^{N_m} A(\sigma_{a},\rho) \right\>_\rho = 
   \left\< A\right\>_\rho,
\label{An1}
\end{equation}
so that  ${\cal O}_1^{\rm est}$ converges to ${\cal O}_1$ 
irrespective of $N_m$:
one could even take $N_m = 1$.\footnote{In practice, one could fix 
$N_m$ in such a way to minimize the variance 
\[
{\rm var}\, {\cal O}_1^{\rm est} = (1/N_s) [
\overline{\< A \>^2} - (\overline{\< A \>})^2 + 
(\overline{\< A^2 \>} - \overline{\< A \>^2})/N_m].
\]
For $\chi$ at fixed $\beta$ this minimization can be done explicitly. 
Indeed, the variance can be related to $U_4$ and $U_{22}$:
$(\hbox{err}[\chi])^2/\chi^2 = [U_{22} + (U_4 - 1)/N_m]/N_s$.
Thus, at the critical point 
$(\hbox{err}[\chi])^2/\chi^2 = (0.148 + 0.648/N_m)/N_s$.
If the work for each disorder configuration 
is proportional to $N_{\rm therm} + N_m$ and $N_{\rm therm} = 300$, 
it is easy to verify that the optimal $N_m$ 
(the one that gives the smallest errors at fixed computational work) 
corresponds to $N_m = 53$.}

 This result could have been guessed directly
from Eq.~(\ref{defOn}), since 
\begin{equation}
{\cal O}_1 = \sum_\rho p(\rho) \pi(\sigma;\rho) A(\sigma,\rho).
\end{equation}
A correct sampling is obtained by determining each time 
a new $\sigma$ and $\rho$ with combined probability $p(\rho) \pi(\sigma;\rho)$. 
Let us now consider $n=2$. We have 
\begin{eqnarray}
\left\< [A]^2\right\>_\rho &=& {1\over N_m^2}
   \left \< \sum_{a=1}^{N_m} \sum_{b=1}^{N_m} 
    A(\sigma_{a},\rho) A(\sigma_{b},\rho) \right\>_\rho
\nonumber \\
   &=& {1\over N_m^2} \left[
   N_m(N_m-1) \left \< A\right\>_\rho^2 + N_m \left \< A^2\right\>_\rho \right] 
\nonumber \\
   &=& \left \< A\right\>_\rho^2 + 
     {1\over N_m} 
    \left( \left \< A^2\right\>_\rho - \left \< A\right\>_\rho^2\right).
\end{eqnarray}
Thus, we obtain for $N_s\to\infty$ at fixed $N_m$
\begin{equation}
{\cal O}_2^{\rm est} \to 
   {\cal O}_2 + {1\over N_m} 
    \left( \overline{\left \< A^2\right\>} - 
           \overline{\left \< A\right\>^2} \right).
\end{equation}
The second term is what is called the {\em bias}. Since in the simulations 
$N_m$ is finite and not too large, this term may give rise to 
systematic deviations 
larger than statistical errors. It is therefore important 
to correct the estimator in such a way to eliminate the bias.

For this purpose
we divide the $N_m$ configurations in $n$ bunches and define the 
sample average over bunch $i$ of length $N_m/n$:
\begin{equation}
[A]_{1/n,i,\rho_\alpha} \equiv  {n\over N_m} \sum_{a=1+(i-1) N_m/n }^{i N_m/n} 
    A(\sigma_{a,\alpha},\rho_\alpha).
\end{equation}
A new estimator of ${\cal O}_n$ is 
\begin{equation}
{\cal O}_n^{\rm unbiased} \equiv  {1\over N_s} \sum_{\alpha=1}^{N_s}
   [A]_{1/n,1,\rho_\alpha} [A]_{1/n,2,\rho_\alpha} \cdots 
   [A]_{1/n,n,\rho_\alpha}.
\label{Onestunb}
\end{equation}
Let us verify that this estimator is unbiased. By repeating the 
arguments presented 
above, for $N_s \to \infty$
the estimator ${\cal O}_n^{\rm unbiased}$ converges to 
\begin{equation}
{\cal O}_n^{\rm unbiased} \to 
  \overline{\left\< [A]_{1/n,1} [A]_{1/n,2} \cdots
   [A]_{1/n,n}\right\>}.
\end{equation}
Because the configurations are assumed to be independent, we have 
\begin{equation}
\hspace{-1cm} 
 \left\< [A]_{1/n,1} [A]_{1/n,2} \cdots
   [A]_{1/n,n}\right\> = 
   \left\< [A]_{1/n,1} \right\> \left\< [A]_{1/n,2} \right\> \cdots 
   \left\<  [A]_{1/n,n}\right\> = \left\<A\right\>^n.
\end{equation}
Thus, for any $n$, irrespective of $N_m$ (one could even take $N_m = n$), 
${\cal O}_n^{\rm unbiased}$ converges to ${\cal O}_n$ as $N_s\to \infty$.

The considerations reported above can be trivially extended to 
disorder averages of products of sample averages. 
Thus, in order to compute $\overline{\< A \> \< B \>}$, we use
\begin{equation}
{1\over 2 N_s} \sum_{\alpha=1}^{N_s} \left(
   [A]_{1/2,1,\rho_\alpha} [B]_{1/2,2,\rho_\alpha}  + 
   [B]_{1/2,1,\rho_\alpha} [A]_{1/2,2,\rho_\alpha}\right) ,
\end{equation}
while for $\overline{\< A \> \< B \> \< C \> }$ we  use
\begin{equation}
{1\over 3! N_s} \sum_{\alpha=1}^{N_s}
   \left\{
   [A]_{1/3,1,\rho_\alpha} [B]_{1/3,2,\rho_\alpha}
   [C]_{1/3,3,\rho_\alpha} +  \hbox{5 permutations}\right\}.
\end{equation}
In the case of $n$ terms, we 
divide the $N_m$ estimates into $n$ parts and then consider
all the $n!$ permutations.

In this paper we extensively use the reweighting technique that requires the 
computation of averages of the form
\begin{equation}
R_{A,B} \equiv  \overline{ \left( {\< A\>  \over \< B\>}\right)},
\end{equation}
where the disorder average should be done after computing the ratio. Indeed,
given a MC run at inverse temperature $\beta$ the mean value of 
an observable $\cal O$ at $\beta + \Delta\beta$ is given by
\begin{equation}
\langle {\cal O} \rangle_{\beta + \Delta\beta} = 
\langle {\cal O} e^{-\Delta\beta {\cal H}}
\rangle_\beta / \langle e^{-\Delta\beta {\cal H}} \rangle_\beta.
\label{reweighting}
\end{equation}
To compute $R_{A,B}$ we consider an estimator of the form
\begin{equation}
R_{A,B}^{\rm est} \equiv  {1\over 2 N_s} \sum_{\alpha=1}^{N_s} \left(
   [A]_{1/2,1,\rho_\alpha} \left\{ {1\over B} \right\}_{1/2,2,\rho_\alpha}  + 
   [A]_{1/2,2,\rho_\alpha} \left\{ {1\over B} \right\}_{1/2,1,\rho_\alpha}  
   \right),
\label{est-RAB}
\end{equation}
where $\{\cdot \}$ should be defined so that 
\begin{equation}
\left\{ {1\over B} \right\} \to {1\over \< B\>} 
\end{equation}
for $N_s\to \infty$ and any $N_m$ 
(the suffix $1/2,i$ has the same meaning as before). 
We have not been able to define
an estimator with this property. We thus use a biased 
estimator, with a bias of order $N_m^{-2}$. Consider 
\begin{equation}
\left\< {1\over [B]_\rho} \right\>_\rho = 
   {1 \over \< B\>_\rho} 
  \left\< \left[ 1 + {1\over N_m} 
   \sum_a \left( { B(\sigma_a,\rho) - \< B\>_\rho \over \< B\>_\rho} \right)
   \right]^{-1} \right \>_\rho .
\end{equation}
Assuming $N_m$ large, we can expand the term in brackets, keeping the 
first nonvanishing term: 
\begin{eqnarray}
&& \hspace{-1.5cm} \left\< {1\over [B]_\rho} \right\>_\rho =
   {1 \over \< B\>_\rho} \left\<
  \left[ 
   1 + {1\over N_m^2} \sum_{ab} 
   \left( { B(\sigma_a,\rho) - \< B\>_\rho \over \< B\>_\rho} \right)
   \left( { B(\sigma_b,\rho) - \< B\>_\rho \over \< B\>_\rho} \right)
   \right] \right\>_\rho + \cdots 
\nonumber \\
 &&=
   {1 \over \< B\>_\rho}
  \left( 1 + {1\over N_m} 
     { \< B^2 \>_\rho - \< B\>_\rho^2 \over \< B\>_\rho^2 } + 
     O(N_m^{-2}) \right).
\end{eqnarray}
Thus, if we define 
\begin{equation}
\left\{ {1\over B} \right\} = 
   {1\over [B]} 
  \left( 1 - {1\over N_m} { [B^2] - [B]^2 \over [B]^2} \right),
\label{def1sub}
\end{equation}
which is such that 
\begin{equation}
\left < \left\{ {1\over B} \right\} \right\> = 
   {1\over \<B\>} (1 + O(N_m^{-2})),
\end{equation}
the estimator (\ref{est-RAB}) converges to $R_{AB}$ with corrections of order 
$N_m^{-2}$. Thus, when using the reweighting technique, $N_m$ is crucial 
and cannot be too small. 

In this paper we also compute derivatives of different observables with 
respect to $\beta$. They can be related to connected correlation function as 
\begin{equation}
\frac{\partial \overline{\langle {\cal O} \rangle}}{\partial \beta} = 
- \overline{ \langle {\cal O} {\cal H} \rangle -
\langle O \rangle \langle {\cal H} \rangle 
}.
\label{derivative}
\end{equation}
Therefore, 
if we apply the reweighting technique, we need to compute terms of the form
\begin{equation}
R_{AB,C} \equiv  \overline{\<A\> \<B\> \over \<C\>^2},
\end{equation}
with $A = {\cal O} e^{-\Delta \beta {\cal H}}$, 
$B = {\cal H} e^{-\Delta \beta {\cal H}}$, and 
$C = e^{-\Delta \beta {\cal H}}$. A possible estimator (this is the one that is used 
in the paper) is 
\begin{equation}
\hspace{-1truecm}  {1\over 4! N_s} \sum_{\alpha=1}^{N_s} \left[ 
   [A]_{1/4,1,\rho_\alpha} [B]_{1/4,2,\rho_\alpha} 
    \left\{ {1\over C} \right\}_{1/4,3,\rho_\alpha}  
    \left\{ {1\over C} \right\}_{1/4,4,\rho_\alpha}  + 
    \hbox {permutations}\right].
\label{ABsuC2}
\end{equation}
The formulae we have derived above rely on two assumptions: 
(i) different configurations obtained with the same disorder $\rho_\alpha$ 
are uncorrelated; (ii) configurations $\sigma$ are extracted with probability 
$\pi(\sigma,\rho)$. None of these two hypothesis is exactly verified in 
practical calculations. Hypothesis (i) is violated because MC simulations 
usually provide correlated sequences of configurations. Correlations change 
some of the conclusions presented above. First, the estimator 
(\ref{Onestunb}) is no longer unbiased, except in the case $n=1$. 
To explain this point, let us consider the specific case $n=2$. 
In the presence of correlations
\begin{equation}
\left\< A(\sigma_{a\alpha},\rho_\alpha) 
        A(\sigma_{b\alpha},\rho_\alpha) \right\>_{\rho_\alpha} = 
    \< A \>_{\rho_\alpha}^2 + (\hbox{var}_{\rho_\alpha} A) C(|a-b|),
\end{equation}
where $\hbox{var}\, A = \<A\>^2 - \<A\>^2$ and $C(t)$ is the 
autocorrelation function. Then
\begin{eqnarray}
&&\hspace{-1cm} 
\left\< [A]_{1/2,1,\rho_\alpha} [A]_{1/2,2,\rho_\alpha}\right\>_{\rho_\alpha}
 \nonumber \\ 
&& = \< A \>_{\rho_\alpha}^2 + 
   {4 \hbox{var}_{\rho_\alpha} A \over N_m^2} \left[
   \sum_{a=1}^{N_m/2} a C(a) + \sum_{a=N_m/2+1}^{N_m-1} 
     (N_m-a) C(a) \right].
\end{eqnarray}
If $C(t)$ decays fast enough the term in brackets is finite 
for $N_m\to \infty$. 
If we further assume $C(t) = \exp(-t/\tau)$, we find that the bias is of order
$(\tau/N_m)^2$. Thus, for $n\ge 2$ the estimator (\ref{Onestunb}) is biased.
Therefore, it is no longer possible to take $N_m$ at will. To avoid the bias, 
$N_m$ should be significantly larger than the autocorrelation time.
Analogously, Eq.~(\ref{def1sub}) is correct only in the absence of
correlations. Otherwise one should define 
\begin{equation}
\left\{ {1\over B} \right\} = 
   {1\over [B]} 
  \left( 1 - {2\tau_{B,{\rm int}}\over N_m} 
     { [B^2] - [B]^2 \over [B]^2} \right),
\label{def1subcorr}
\end{equation}
where $\tau_{B,{\rm int}}$ is the integrated autocorrelation time associated 
with $B$. In the present work the MC algorithm is very efficient, so that
our measurements should be nearly independent. Thus, we have always used 
Eq.~(\ref{def1sub}).

The second assumption that we have made is that 
configurations $\sigma_{a,\alpha}$ are 
extracted with probability $\pi(\sigma,\rho_\alpha)$. In MC calculations 
this is never the case: configurations are obtained by using a Markov chain 
that starts from a nonequilibrium configuration. Therefore $\sigma_{a,\alpha}$
are generated by using a distribution that converges 
to $\pi(\sigma,\rho_\alpha)$
only asymptotically. This is the so-called inizialization bias. 
This bias, which is of 
order $N_m^{-1}$,\footnote{More precisely, if we define 
\[
[A]_{N_{\rm th}} = {1\over N_m} \sum_{k = 1 + N_{\rm th}}^{N_m + N_{\rm th}}
  A(\sigma_k),
\]
then $N_m (\< [A]_{N_{\rm th}}\>_{MC} - \<A\>)$ converges to a nonvanishing 
constant $K$ as $N_m\to\infty$ (here $\<\cdot\>_{MC}$ is an average over all 
Markov chains that start at $i=0$ from an arbitrary configuration). 
The constant $K$ decreases as $N_{\rm th}$ increases, with a rate 
controlled by the exponential autocorrelation time.}
cannot be avoided. 
However, by discarding a sufficiently large
number of initial configurations one can reduce it arbitrarily. 
Note that, as $N_s$ 
is increased, either $N_m$ or the number of discarded configurations 
should be increased too.

\begin{figure}[tb]
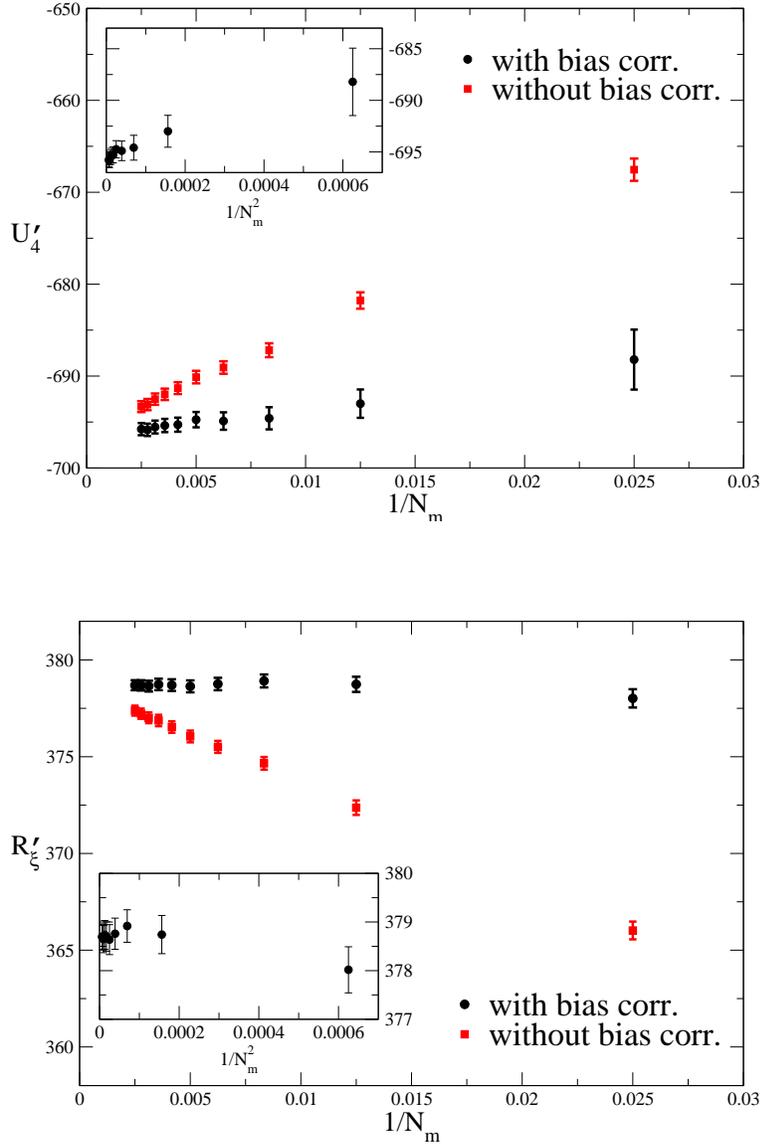

\vspace{10mm}
\centerline{\psfig{width=10truecm,angle=0,file=dbinder_inverso.eps}}
\vspace{12mm}
\centerline{\psfig{width=10truecm,angle=0,file=dxi_inverso.eps}}
\vspace{2mm}
\caption{$U'_4$ and $R'_\xi$ for a run of the RSIM model at $L=64$, $p=0.8$, 
$N_s=60000$, 
$\beta_{\rm run}=0.285742$, reweighted at $R_\xi$ = 0.5943 as a 
function of $1/N_m$. In the insets we show the results for the 
bias-corrected estimates versus $1/N_m^2$. We report data with
$N_m = 40,80,120,\ldots,400$.
}
\label{figNm}
\end{figure}

Of course, the practically interesting question is whether, with the values of 
$N_m$ and $N_s$ used in the MC simulations, bias corrections are relevant 
or not. 
For this purpose, in Fig. \ref{figNm} we compare, for a specific run 
of the RSIM at $p=0.8$, the value of the derivatives $U'_4$, $R'_\xi$, with and 
without bias correction, as a function of $N_m$; 
data obtained for $\beta = 0.2857420$ are reweighted to have $R_\xi=0.5943$
[this corresponds to $\beta_f = 0.2857478(18)$].
The average without bias correction has 
been estimated by using 
\begin{equation}
 {1\over N_s} \sum_{\alpha=1}^{N_s} 
   {[A]_{\rho_\alpha} [B]_{\rho_\alpha} 
    \over [C]^2_{\rho_\alpha} } 
\end{equation}
while the bias-corrected estimate corresponds to Eq.~(\ref{ABsuC2}).
From the figure, we see that the biased estimate shows a systematic drift,
which, as expected, is linear in $1/N_m$. 
The bias-corrected estimate 
is instead essentially flat; deviations are observed only for $U_4'$ for 
$N_m \lesssim 100$. As expected, they apparently decrease as $1/N_m^2$.
Note, that even for $N_m = 400$ the biased estimate differs within 
error bars from the bias-corrected one. Thus, with the precision of our 
calculations, the bias correction is essential for 
both $R_\xi'$ and $U_4'$.

We would like to point out that the expressions obtained here are 
somewhat different
from those proposed in Ref.~\cite{BFMMPR-98-b}. Only for 
$\overline{\<A\> \<B\>}$
are they identical.  We report here the expressions for $R_{AB}$ and 
$R_{AB,C}$:
\begin{eqnarray}
R_{A,B}^{\rm est} &= &
   {2\over N_s} \sum_{\alpha=1}^{N_s} \left[ 
   {[A]\over [B]} - {[A]_{1/2,1} \over 4 [B]_{1/2,1}} - 
       {[A]_{1/2,2} \over 4 [B]_{1/2,2}} \right], \nonumber \\
R_{AB,C}^{\rm est} &= &
   {2\over N_s} \sum_{\alpha=1}^{N_s} \left[ 
   {[A][B]\over [C]^2} - {[A]_{1/2,1} [B]_{1/2,1} \over 4 [C]_{1/2,1}^2} - 
   {[A]_{1/2,2} [B]_{1/2,2} \over 4 [C]_{1/2,2}^2} \right] .
\end{eqnarray}
The idea behind these formulas is the following. Consider ${\cal O}$ 
which is an arbitrary function of thermal averages. To compute 
$\overline{\cal O}$, consider an arbitrary estimator ${\cal O}^{\rm est}$ of 
$\overline{\cal O}$. For $N_s\to \infty$ at fixed $N_m$, ${\cal O}^{\rm est}$
converges to
\begin{equation}
    \overline{\cal O} + {a\over N_m} + O(1/N_m^2).
\label{as-O}
\end{equation}
A better estimator, without the $1/N_m$ correction, is obtained by
considering
\begin{equation}
{\cal O}^{\rm est,unb} = 2 {\cal O}^{\rm est} - 
   {1\over 2} {\cal O}^{\rm est}_{1/2,1} - 
   {1\over 2} {\cal O}^{\rm est}_{1/2,2}.
\label{OParisi}
\end{equation}
Here ${\cal O}^{\rm est}$ is determined by using all $N_m$ measures, 
while ${\cal O}^{\rm est}_{1/2,1}$ and ${\cal O}^{\rm est}_{1/2,2}$
are computed by using the first half and the second half of the measures,
respectively. It is easy to see, by substituting the behavior 
(\ref{as-O}) in Eq.~(\ref{OParisi}), that the new estimator
has no corrections of order $1/N_m$.

        \end{document}